\documentclass[acmsmall,nonacm,screen]{acmart}

\usepackage[utf8]{inputenc}
\usepackage[english]{babel}
\usepackage{xspace}
\usepackage[caption=false]{subfig}
\usepackage{tikz}
\usetikzlibrary{patterns}
\usepackage{amsthm}

\newcommand{\system}[0]{\textsc{Spider}\xspace}
\newcommand{\channel}[0]{IRMC\xspace}
\newcommand{\channela}[0]{\mbox{\channel-RC}\xspace}
\newcommand{\channelb}[0]{\mbox{\channel-SC}\xspace}
\newcommand{\msg}[2]{\ensuremath{\langle \textsc{#1}, #2 \rangle}\xspace}
\newcommand{\smsg}[3]{\ensuremath{\langle \textsc{#1}, #2 \rangle_{#3}}\xspace}
\newcommand{\bft}[0]{BFT\xspace}
\newcommand{\hft}[0]{HFT\xspace}
\newcommand{\bftwv}[0]{BFT-WV\xspace}
\newcommand{\systemze}[0]{\mbox{\system-0E}\xspace}
\newcommand{\systemoe}[0]{\mbox{\system-1E}\xspace}

\newcommand{\headline}[1]{\vspace{1mm plus .5mm minus .5mm}\noindent\textbf{\textit{#1.}}~}

\usepackage{listings}
\usepackage{subscript}

\font\lsttt=rm-lmtl10 scaled 820
\font\lstbtt=rm-lmtk10 scaled 820

\newcommand{\commentsize}{\fontsize{8pt}{0pt}\selectfont}
\newcommand{\pseudocode}{\commentsize\normalfont}

\begin{document}

\title{\system: A BFT Architecture for Geo-Replicated Cloud Services}

\author{Michael Eischer}
\affiliation{%
	\institution{Friedrich-Alexander-Universität Erlangen-Nürnberg}
	\city{Erlangen}
	\country{Germany}
}

\author{Tobias Distler}
\affiliation{%
	\institution{Friedrich-Alexander-Universität Erlangen-Nürnberg}
	\city{Erlangen}
	\country{Germany}
}

\makeatletter
\let\@authorsaddresses\@empty
\makeatother

\thanks{This paper extends on earlier work that was presented at the Middleware 2020 conference~\cite{eischer20resilient} and along with a formal safety and liveness proof has also been published at the Computing Research Repository~\cite{eischer20resilient-extended}. The main additions include a refined and more detailed problem statement~(Section~\ref{sec:problem}), an overview of the architecture's building blocks to emphasize the approach's modularity~(Section~\ref{sec:building-blocks}), a description of a protocol extension for providing sequential consistency~(Section~\ref{sec:protocol}), a presentation of several optimizations improving the system's effectiveness and efficiency~(Section~\ref{sec:optimizations}), and new evaluation results from an extended experimental analysis of different channel implementations~(Section~\ref{sec:evaluation-channels}). This work was partially supported by the Deutsche Forschungsgemeinschaft (DFG, German Research Foundation) -- 157267460~(``REFIT'')}

\begin{abstract}

Traditionally, Byzantine fault tolerance~(BFT) in geo-replicated systems is achieved by executing complex agreement protocols over large-distance communication links, and therefore typically incurs high response times. In this paper we address this problem with \system, a resilient and modular BFT replication architecture for geo-distributed systems that leverages characteristic features of today's public-cloud infrastructures to minimize both complexity as well as latency. \system is composed of multiple largely independent replica groups that each are distributed across different availability zones of their respective cloud region. This design offers the possibility to provide low response times by placing replica groups in close geographic distance to clients, while at the same time enabling intra-group communication over short-distance links. To handle the interaction between groups that is necessary for strong consistency, \system uses a novel message-channel abstraction with first-in-first-out semantics and built-in flow control that greatly simplifies system~design.

\end{abstract}

\maketitle

\section{Introduction}

Byzantine fault-tolerant~(BFT) protocols enable a system to withstand arbitrary faults and consequently have been used to increase the resilience of a wide spectrum of critical applications such as key-value stores~\mbox{\cite{padilha13augustus,padilha16callinicos,li16sarek,eischer19deterministic}}, SCADA systems~\cite{nogueira18challenges,babay18network,babay19deploying}, firewalls~\cite{bessani08crutial,garcia16sieveq}, coordination services~\cite{clement09upright,kapitza12cheapbft,behl15consensus,distler16resource,eischer19scalable}, and permissioned blockchains~\cite{sousa18byzantine,gueta19sbft}. To provide their high degree of fault tolerance, BFT protocols replicate the state of an application across a set of servers and rely on a leader-based consensus algorithm to keep these replicas consistent. This task requires several subprotocols~(e.g.,~for leader election, checkpointing, state transfer) and multiple phases of message exchange between replicas~\cite{castro99practical}.

Unfortunately, this complexity makes it inherently difficult to achieve low latency in use cases in which the clients of an application are scattered across various geo\-graphic locations. For example, placing replicas in close proximity to each other may reduce the latency of strongly consistent requests whose execution must be coordinated by the consensus protocol between replicas. However, with replicas being located farther apart from clients this strategy also increases the response times of requests such as weakly consistent reads that do not need to be agreed on and only involve direct interaction between clients and replicas. In contrast, co-locating replicas with clients has the inverse effect of speeding up client--replica communication but adding a significant performance overhead to the agreement protocol.

Existing approaches for BFT wide-area replication aim at minimizing this overhead by (1)~applying weighted-voting schemes to reduce the quorum sizes needed to complete consensus~\cite{sousa15separating,berger19resilient}, (2)~rotating the leader role among replicas to shorten the path necessary to insert a request into the agreement protocol~\mbox{\cite{veronese09spin,mao09towards,veronese10ebawa}}, or (3)~relying on a two-level system design that deploys an entire BFT replica cluster at each client site in order to be able to use crash-tolerant replication between sites~\cite{amir10steward,amir07customizable}. In all these cases, BFT systems still need to run complex consensus-based replication protocols over wide-area links which not only results in response-time overhead but also makes it difficult to dynamically introduce new replica sites, for example, to serve clients at new locations.

In this paper we address these problems with \system, a cloud-based BFT system architecture for geo-replicated services that models a system as a collection of loosely coupled replica groups that are deployed in different regions. Separating agreement from execution~\cite{yin03separating}, one of the groups (``\emph{agreement group}'') establishes an order on all requests with strong consistency demands while all other groups~(``\emph{execution groups}'') are responsible for communicating with clients and processing requests. In contrast to existing approaches, \system does not require complex wide-area protocols but instead handles tasks such as consensus, leader election, and checkpointing within a group and over short-distance links. To make this possible while still offering resilience against replica failures, \system leverages the design of today's cloud infrastructures~\cite{ec2-regions,azure-regions,gce-regions} and places the replicas of a group in different availability zones of the same region; availability zones are hosted by data centers at distinct sites and specifically engineered to represent different fault domains.

In particular we make five contributions in this paper: (1)~We present the \system architecture and discuss how it achieves low latency for weakly consistent reads by placing execution groups close to clients, while at the same time minimizing agreement response times for strongly consistent reads and writes. (2)~We show how to design \system in a modular way so that execution groups do not depend on internals of the agreement group~(e.g.,~a specific consensus protocol). As an additional benefit, the modularity also makes it straightforward to add/remove execution groups at runtime. (3)~We introduce a wide-area BFT flow-control mechanism that exploits the special characteristics of \system to minimize complexity. Our approach is based on a simple message-channel abstraction that handles the inter-regional communication between two replica groups and prevents one group from overwhelming the other. (4)~We present several optimizations to reduce the overhead for authenticating messages, thereby enabling \system to not only provide low latency but also sustain high throughput. (5)~We experimentally evaluate \system in a public-cloud environment~(Amazon EC2) in comparison to the state of the art in BFT wide-area replication.

\vspace{-.3mm}

\section{System Model}
\label{sec:model}

Our work focuses on stateful applications with strong reliability requirements whose clients are scattered across different geographic locations, potentially all over the globe. To access the application a client submits a request to the server side and waits for a result. We assume that both clients and servers can be subject to Byzantine faults. As a consequence, nodes~(i.e.,~clients and servers) do not trust each other and do not make irreversible decisions based on the input provided by a single other node alone. For example, to tolerate up to $f$~faulty servers, a client only accepts a result after it has obtained at least $f+1$~matching replies from different servers.

Besides service availability and correctness in the presence of failures, low latency is a primary concern in our target systems. Achieving this goal while keeping the states of servers consistent is inherently difficult in use cases in which clients are geographically dispersed. The problem is further complicated by the fact that we assume that the locations from which clients access the application may change over time, typically as a result of the global day/night cycle. To continuously provide low latency under such conditions, a system must offer some kind of reconfiguration mechanism enabling an adaptation to varying workloads. One possibility to achieve this, for example, is to dynamically include additional servers that are located closer to newly started clients.

\vspace{-.3mm}

\section{Background}
\label{sec:background}

In the following, we elaborate on the problems associated with Byzantine fault tolerance in geo-distributed systems and discuss existing approaches to solve them.


\begin{figure}
 	\subfloat[PBFT~\cite{castro99practical}\label{fig:pbft}] {
		\includegraphics{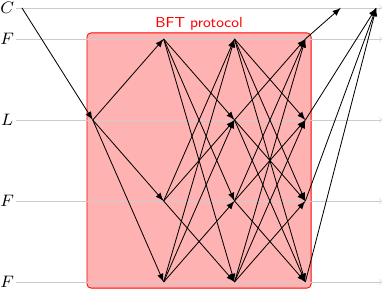}
	}
	\hspace{2mm}
	\subfloat[Steward~\cite{amir10steward}\label{fig:steward}] {
		\includegraphics{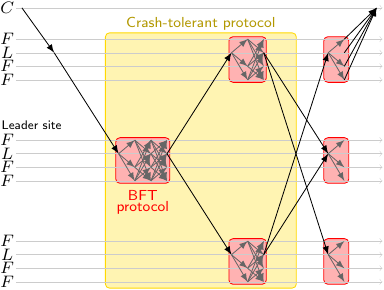}
	}
	\caption{BFT geo-replication architectures connecting a client~(C) with leader~(L) and follower~(F) replicas.}
\end{figure}

\headline{BFT in Wide-Area Environments}
%
The straightforward approach to offer resilience against arbitrary failures is to rely on a BFT replication protocol, for example PBFT~\cite{castro99practical}. As illustrated in Figure~\ref{fig:pbft}, PBFT requires at least $3f+1$~replicas to tolerate $f$~failures. To keep the application state consistent across replicas, PBFT ensures that replicas run an agreement protocol to decide in which order to process client requests. For this purpose, PBFT elects one of the replicas as leader~(marked $L$ in Figure~\ref{fig:pbft}) while all other replicas assume the roles of followers~($F$). Having received a new request, the leader is responsible for initiating the agreement process, which then involves multiple message exchanges between replicas. To deal with scenarios where a faulty leader does not behave according to specification, for example by ignoring a request, PBFT provides a mechanism that enables followers to depose the leader and appoint a new one. Once the agreement process is complete, all non-faulty replicas execute the request and send the result to the client, thereby enabling the client to validate the result by comparison.

Using BFT protocols such as PBFT to build resilient systems is effective but has several disadvantages in the context of geo-replication: (1)~With replicas being distributed across different geographic sites, the entire BFT~protocol needs to be executed over wide-area links, which often results in high response times. Note that this is not only true with regard to the task of agreeing on requests during normal operation, but for example also for electing a new leader as part of fault handling. (2)~Due to the fact that all requests must flow through the leader, the geographic location of the leader, and in particular its position relative to the majority of followers, usually has a significant influence on latency~\mbox{\cite{sousa15separating,eischer18latency}}. Consequently, a leader switch may decisively change a system's performance characteristics, requiring clients to deal with the associated latency volatility. (3)~Consisting of only $3f+1$~replicas, for traditional BFT systems it is inherently difficult to select suitable replica locations in cases where a large and varying number of clients are scattered across the globe. Ideally, replicas would be placed both in close distance to each other (to speed up agreement) as well as in close distance to clients (to minimize the transmission time of requests and results). For systems with just a few replicas but many clients meeting this requirement is essentially impossible.

\headline{Weighted Voting}
%
By assigning different weights on the votes which replicas have within the consensus protocol~\cite{sousa15separating,berger19resilient} it is feasible to introduce additional replicas while keeping response times low or even reducing them in a geo-replicated setting. Unfortunately, this comes at the cost of an increased number of messages exchanged between replicas, which can be prohibitively expensive in public-cloud settings as providers typically charge extra for wide-area traffic.

\headline{Leader Rotation}
%
Different authors have proposed to improve performance by rotating the leader role among replicas, following the idea of enabling each client to submit requests to its nearest replica~\mbox{\cite{veronese09spin,mao09towards,veronese10ebawa}}. Results from an extensive experimental evaluation by Sousa~et~al.~\cite{sousa15separating}, however, showed that in practice this approach does not provide significant benefits compared with appointing a fixed leader at a well-connected site. Besides, leader rotation still requires the execution of a complex protocol over wide-area links.

\vspace{2mm}

\headline{Hierarchical System Architecture}
%
To increase the scalability of BFT systems in wide-area settings, Amir et al. presented a hierarchical architecture as part of Steward~\cite{amir10steward}. As shown in Figure~\ref{fig:steward}, instead of hosting a single replica, each site in Steward comprises a cluster of replicas that run a site-local BFT agreement protocol. A key benefit of this approach is the fact that, although individual replicas still may be subject to Byzantine faults, an entire cluster can be assumed to only fail by crashing. This property at the local level enables Steward to rely on a crash-tolerant agreement protocol at the global level~(i.e.,~between sites), which compared with traditional BFT systems requires fewer phases and fewer message transmissions over wide-area links.

The efficiency enhancements made possible by its architecture enable Steward to improve performance, however, they come at the cost of an increased overall complexity that stems from the need to maintain replication protocols at two levels: within each site as well as between sites. Designing and implementing such protocols in isolation is already a non-trivial task, additionally guaranteeing a correct interplay between them is even more challenging. To ensure liveness Steward, for example, requires timeouts at different levels to be carefully coordinated~\cite{amir10steward}. Amir et al. addressed these problems in a subsequent work~\cite{amir07customizable}, which in this paper we refer to as CFT-WAR. In contrast to Steward, in CFT-WAR each step of the wide-area protocol~(e.g.,~Paxos~\cite{lamport98part}) is handled by a full-fledged multiphase consensus protocol at each site~(e.g.,~PBFT). As a main advantage, this approach disentangles the protocols used for wide-area and site-internal replication. On the downside, it introduces additional overhead that in general prevents CFT-WAR from achieving response times as low as Steward's when providing the same degree of fault tolerance~\cite{amir07customizable}. Furthermore, due to performing agreement at two levels CFT-WAR still needs to run multiple subprotocols for tasks such as leader election, one at each level.
A set of additional subprotocols would be required to support the dynamic addition/removal of individual replicas or entire sites in a hierarchical system architecture, thereby further increasing complexity. To our knowledge, the ability to adjust to varying workload conditions was not a design goal of Steward and CFT-WAR, which is why the systems do not offer mechanisms for changing their composition at runtime.

\vspace{2mm}

\section{Problem Statement}
\label{sec:problem}

Our analysis in Section~\ref{sec:background} shows that applying existing approaches to provide BFT in a cloud-based geo-replicated environment is possible, for example with regard to safety, but cumbersome due to the associated high complexity and the lack of effective means to react to changing workloads. This observation led us to ask whether these problems can be circumvented by a BFT system architecture that is specifically tailored to the characteristics of today's cloud infrastructures. In particular, we aim for an architecture that provides efficiency, modularity, consistency, and adaptability.

\vspace{2mm}

\headline{Low End-to-End Latency}
%
To minimize response times during both normal-case operation as well as fault handling, a system architecture in the ideal case does not require the execution of complex protocols over wide-area links. Instead, tasks involving multiple phases of message exchange between replicas, such as the agreement on requests, should be handled by replicas that are located in comparably close distance to each other.

\newpage

\emph{Our approach:} Exploiting the design of today's public clouds, we propose a resilient system architecture that organizes a BFT system as a collection of replica groups that each serve a specific purpose, which is either the agreement or the execution of requests. Since each of these replica groups runs distributed across different cloud-provided fault domains within the same geographic region, interactions between replicas of the same group are solely handled via short-distance links. Overall, this enables our architecture to reduce the wide-area overhead for agreed requests to two communication steps, which is the minimum for any (crash-tolerant or Byzantine fault-tolerant) system in the geo-distributed environments we target with our work.

\headline{Modular Structure}
%
Supporting a variety of use cases with different requirements is difficult if the protocols responsible for agreement and execution are hard-wired into the system architecture. To address this issue, we join other authors~\cite{amir07customizable} in aiming for an architecture that can be combined with different consensus protocols depending on the specific demands of an application.

\emph{Our approach:} The BFT system architecture we have developed comprises a hierarchical structure that offers modularity based on two key components. First, the agreement protocol used to reach consensus on a total processing order for client requests is treated as a black box and therefore can be tailored to the individual characteristics of the replicated service. The interface required from this black box is well-defined and specified in such a way that several existing agreement protocols can be integrated with the architecture. Second, to achieve composability, we present a novel channel abstraction that serves as connecting piece between replica groups in our system. The channel abstraction is able to safely transfer information from one replica group to another, and thanks to its generic design the abstraction can be used to handle all inter-group communication~(in both directions) without introducing further subprotocols.

\headline{Multi-Level Consistency}
%
With the overhead of achieving consistency usually increasing for strong guarantees such as linearizability~\cite{herlihy90linearizability}, wide-area architectures for replicated systems should enable clients to benefit from latency improvements that are made possible by exploiting weaker forms of consistency~\cite{terry13consistency}. For example, if a client application can cope with reading old, potentially outdated values as results for its operations, then overall response times may be reduced by allowing the corresponding requests to bypass the agreement protocol~\cite{amir10steward}.

\emph{Our approach:} In our BFT system architecture, clients have the flexibility to choose between multiple consistency levels on a per-request basis, possibly selecting different guarantees for different requests. While writes are always linearizable, the guarantees for individual reads may range from linearizability to prefix consistency~\cite{viotti16consistency}. As a key benefit, the latter for example can be directly provided by a client's closest replica group, therefore typically requiring no wide-area communication throughout the entire read operation.

\headline{Dynamic Adaptability}
%
One major strength of public clouds is to quickly provide resources on demand and at various geographic locations. A BFT architecture should be able to leverage this feature for hosting replicas in the proximity of clients to reduce the latency with which clients access the replicated service. Specifically, if new clients are started at other sites, there should be a lightweight mechanism for dynamically adding new replicas. The same applies to means for removing replicas that are no longer of benefit as the clients in their vicinity have been shut down.

\emph{Our approach:} Separating agreement and execution into distinct types of replica groups, our modularized system architecture makes it straightforward to efficiently support clients all over the globe. Thanks to a loose coupling between replica groups, new groups can be quickly added to or removed from the system in order to adjust to changes in client numbers and locations. To assist new clients in connecting to the system, we provide means that enable clients to determine the addresses of replica groups that reside close to their own geographic positions.

\section{The S{\footnotesize{}pider} Architecture}
\label{sec:architecture}

\begin{figure}[b!]
	\vspace{-1mm}
	\includegraphics{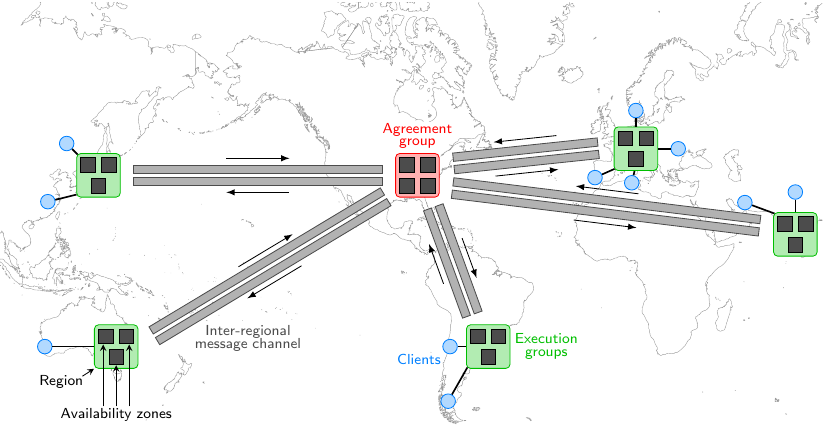}
	\vspace{-6mm}
	\caption{\system system architecture}
	\label{fig:architecture}
\end{figure}

This section presents \system, a cloud-based BFT system architecture that targets wide-area use cases and therefore is distributed across multiple geographic sites. For this purpose, \system leverages the common organizational structure of state-of-the-art cloud infrastructures such as Amazon EC2~\cite{ec2-regions}, Microsoft Azure~\cite{azure-regions}, or Google Compute Engine~\cite{gce-regions} by grouping sites into \emph{regions}, as shown in Figure~\ref{fig:architecture}. The sites within a region typically are several tens of kilometers apart from each other and represent separate fault domains, commonly referred to as \emph{availability zones}. In addition to constructing the data centers at distinct geographic locations, cloud providers also ensure that data centers in different availability zones are equipped with dedicated power supply systems and network links to minimize the probability of dependent failures. For the \system system architecture, availability zones play an important role as they allow us to place replicas in separate fault domains and still enable them to interact over short-distance links with comparably low latency.


\headline{Replica Groups}
%
Relying on this setting, \system is composed of multiple loosely coupled replica groups, each being distributed across different availability zones of a specific region. One of the replica groups in the system, the \emph{agreement group}, is responsible for establishing a global total order on incoming requests. The size of this group depends on the protocol it uses for consensus. Running PBFT~\cite{castro99practical}, for example, the agreement group consists of $N_a=3f_a+1$~replicas and is able to tolerate $f_a$~Byzantine faults. All other replica groups in the system, the \emph{execution groups}, host the application logic, process the ordered requests, and handle the communication with clients. Each of these groups comprises $N_e=2f_e+1$~replicas and tolerates at most $f_e$~Byzantine faults.
The level of fault tolerance provided by the agreement group and the execution groups may be selected independently. Supporting multiple execution groups enables \system to scale throughput by adding/removing groups and to minimize latency by placing groups in the vicinity of clients.

\headline{Execution-Replica Registry}
%
\system contains an execution-replica registry to provide clients with information on the locations and addresses of active replicas. The registry is a BFT service that is hosted and maintained by the agreement group. Its contents are updated by agreement replicas whenever the composition of the system changes~(see Section~\ref{sec:adaptability}).

\headline{Low-Latency BFT Replication}
%
In contrast to existing ap\-proach\-es~(see Section~\ref{sec:background}), \system does not run a full-fledged and complex replication protocol over long-distance links. Instead, all non-trivial tasks~(e.g.,~reaching consensus on requests) are carried out within a replica group using low-latency intra-regional connections. Following this design principle, \system handles requests by forwarding them along a chain of stages represented by different replica groups. Specifically, clients submit their requests to their nearest execution group, which in turn forwards the request to the agreement group for ordering. Once this step is complete, the agreement group instructs all execution groups to process the ordered request.
This ensures that execution-group states remain consistent without requiring the execution groups to reach consensus themselves. Having processed the request, the replicas of the execution group the client is connected to return the result. As each execution group comprises $2f_e+1$~replicas, clients are able to verify the correctness of a result solely based on the replies they receive from their local execution group.

With all communication-intensive steps being performed over intra-regional links, inter-regional links in \system are only responsible for forwarding the outputs of one stage to the replica group(s) constituting the next stage. In particular, this approach has the following benefits: (1)~It greatly simplifies the interaction of replicas over long-distance connections. (2)~It enables a modular design that allows different deployments to rely on different agreement protocols without the need to modify the implementation of execution replicas. (3)~As we show in Section~\ref{sec:channels}, it allows \system to use the same abstraction, a reliable message channel, for all inter-regional links, thereby facilitating system implementation.

\headline{Practical Considerations}
%
As of this writing, all major public clouds offer several regions with at least three availability zones~(Amazon EC2:~33, Microsoft Azure:~30, Google Compute Engine:~40) and therefore support the world-wide deployment of \system execution groups which tolerate one faulty replica. In addition, Amazon~(Virginia, Oregon, Seoul, Tokyo) and Google~(Iowa) also already operate regions with four or more availability zones, which consequently are candidates for hosting \system's agreement group. With public cloud infrastructures still being expanded, new regions and availability zones are added every year, increasing the deployment options for \system. Besides, to further improve the resilience of \system, agreement and execution replicas may be distributed across different clouds, thereby reducing the dependence on a single provider~\cite{bessani13depsky,abu-libdeh10racs}. As there are several regions hosting data centers and availability zones of multiple providers~(e.g.,~Europe, North America, South America, India, Asia, and Australia), this approach also makes it possible to deploy larger agreement and execution groups that tolerate $f_a>1$ and $f_e>1$ replica failures, respectively.

Representing distinct fault and upgrade domains, availability zones are designed to enable uninterrupted execution of services that are replicated within the same region. Despite the efforts undertaken by providers, in the past there have been rare incidents where problems in one availability zone caused temporary availability issues in other zones belonging to the same region~\cite{aws11incident}. In \system, if more than $f_a$~agreement replicas are unresponsive, the agreement group temporarily cannot order new requests until the replicas become available again. However, as we detail in Section~\ref{sec:protocol}, in such cases \system is still able to process weakly consistent read requests as these operations are handled within a client's local execution group. On the other hand, if more than $f_e$~replicas of the same execution group become unavailable, affected clients can temporarily switch to a different execution group and continue to use the service.

\section{Building Blocks}
\label{sec:building-blocks}

This section discusses the main components that, when assembled together, form \system's modular system architecture. Specifically, this includes (1)~the consensus protocol used to totally order requests in the agreement group, (2)~the message-channel abstraction connecting the agreement group with all execution groups, as well as (3)~the application logic running in each execution group. A summary of the components' interfaces is presented in Figure~\ref{fig:interfaces} and referred to throughout this section. For ease of presentation, for the most part we describe the interfaces using synchronous~(i.e.,~potentially blocking) methods, however implementations relying on a higher extent of asynchronous interactions~(e.g.,~non-blocking method calls combined with event-triggered callbacks) are also~possible.


\begin{figure}[b]
	\hrule\vspace{.25mm}\begin{center}\lstbtt{Agreement}\end{center}\hrule\vspace{1.5mm}
	\begin{lstlisting}
%\lstbtt{interface}% Agreement {
	%\textsc{Void}% %\lstbtt{order}%(%\textsc{ClientID}% c, %\textsc{ClientCtr}% tc, %\textsc{Request}% r);
	%\lstbtt{callback}% %\textsc{Void}% %\lstbtt{ordered}%(%\textsc{SeqNr}% s, %\textsc{Request}% r);
	%\textsc{Void}% %\lstbtt{collect\_garbage}%(%\textsc{SeqNr}% s, %\textsc{ClientCtr}%[] ts);
}%\vspace{2mm}\hrule\vspace{.25mm}\begin{center}\lstbtt{Inter-Regional Message Channel}\end{center}\hrule\vspace{2mm}%
%\lstbtt{interface}% IRMC_Sender {
	%\textsc{Void}% %\lstbtt{send}%(%\textsc{Subchannel}% sc, %\textsc{Position}% p, %\textsc{Message}% m);
	%\textsc{Void}% %\lstbtt{move\_window}%(%\textsc{Subchannel}% sc, %\textsc{Position}% p);
}

%\lstbtt{interface}% IRMC_Receiver {
	%\textsc{Message}% %\lstbtt{receive}%(%\textsc{Subchannel}% sc, %\textsc{Position}% p);
	%\textsc{Void}% %\lstbtt{move\_window}%(%\textsc{Subchannel}% sc, %\textsc{Position}% p);
}%\vspace{2mm}\hrule\vspace{.25mm}\begin{center}\lstbtt{Application}\end{center}\hrule\vspace{2mm}%
%\lstbtt{interface}% Application {
	%\textsc{Result}% %\lstbtt{execute}%(%\textsc{Request}% r);
	%\textsc{Boolean}% %\lstbtt{is\_read\_only}%(%\textsc{Request}% r);
	%\textsc{State}% %\lstbtt{snapshot}%();
	%\textsc{Void}% %\lstbtt{apply}%(%\textsc{State}% st);
}%\vspace{2mm}\hrule%
\end{lstlisting}
	\caption{Interfaces of \system's main building blocks}
	\label{fig:interfaces}
\end{figure}

\subsection{Agreement Protocol Black Box}
\label{sec:agreement}

The main task of \system's agreement group is to establish a total order in which client requests later need to be processed by all execution groups. To increase flexibility, we decided not to hard-wire a specific ordering protocol into our architecture and instead enabled agreement replicas to treat the consensus process as a black box. The interface and requirements of this black box are designed in such a way that many agreement protocols~\cite{castro99practical,bessani14state,amir10prime,clement09making,kotla09zyzzyva} are able to implement it, either directly or with only minor adaptations.

\headline{Request Ordering}
%
Similar to existing replication protocols~\cite{castro99practical,bessani14state}, each client request in \system is uniquely identified by a tuple $\langle c, t_c \rangle$ containing the id~$c$ of the client that submitted the request as well as a monotonically increasing client-specific counter value~$t_c$ representing a unique logical timestamp. When an agreement replica provides this information for a new request to the agreement protocol, the protocol is responsible for assigning the request with a sequence number that remains stable even in case of faults~(\texttt{order()}, see Figure~\ref{fig:interfaces}). To publish the outcome of the consensus process~(i.e.,~a committed sequence number and its associated value), the agreement protocol black box notifies its local agreement replica by invoking a callback method \texttt{ordered()} in increasing order of sequence numbers. As committed value the agreement protocol in \texttt{ordered()} may either provide a regular client request or a special no-op request. The latter allows \system to employ agreement protocols that use no-ops as fillers during view change~\cite{castro99practical}. No-ops have no effect on the application state since they are filtered out prior to execution.

\headline{Required Properties}
%
In order to be a candidate for \system's agreement protocol black box, a consensus algorithm needs to tolerate up to $f_a$ Byzantine faults using $N_a$~agreement replicas. Furthermore, an agreement protocol must satisfy the following three properties:

\begin{itemize}
	\item \textbf{Safety}~~The protocol must guarantee that the request delivered for a sequence number is the same at all correct replicas. As a main consequence, this property allows correct agreement replicas to consistently publish the consensus outcome to the rest of the system.
	\item \textbf{Liveness}~~Requests submitted by at least $N_a-f_a$ correct agreement replicas must be eventually assigned a stable sequence number on at least $f_a+1$ correct agreement replicas. Agreement protocols commonly meet this requirement by monitoring an incoming client request with a timer and replacing the leader replica in case the timer expires without the request being committed. Requests that completed the consensus process must be delivered via \texttt{ordered()} callback unless they are affected by a garbage-collection hint~(see~below).
	\item \textbf{Validity}~~The resulting total order put out by the agreement protocol must only be composed of well-formed, correctly authenticated client requests and no-op requests.
\end{itemize}

\headline{Garbage Collection}
%
To reclaim memory, agreement replicas in \system discard information about ordered requests after they have been safely forwarded to execution groups. In an effort to enable the consensus protocol to do the same, replicas on such occasions also provide garbage-collection hints to the agreement protocol black box by invoking a \texttt{collect\_garbage()} method. Once the method is called for a given sequence number, all slots with earlier sequence numbers may be garbage collected and no longer need to be delivered by the \texttt{ordered()} callback; the also provided array containing the latest client counters can be used to stop timers for requests that were garbage collected. As a potential side effect, the garbage collection can cause the consensus instances for some sequence numbers to be skipped, for example when a replica applies a newer checkpoint.

\headline{Flow Control}
%
\system supports the implementation of agreement replicas with bounded space provided that the agreement protocol black box limits the number of requests queued for ordering as well as the number of requests waiting to be delivered. Such a flow-control mechanism commonly exists in agreement protocols. PBFT~\cite{castro99practical}, for example, implements this feature by using high and low watermarks to limit the number of currently active sequence numbers. Similarly, BFT-SMaRt~\cite{bessani14state} processes sequence numbers one after another and combines this concept with a bound on the number of requests waiting for execution.

\subsection{Inter-Regional Message Channels}
\label{sec:channels}

To support a modular design, we use an abstraction to handle all interaction between replica groups in \system: the \emph{inter-regional message channel~(\channel)}. Specifically, \channel{}s are responsible for forwarding messages from a group of sender replicas in one region to a group of receiver replicas in another region. Conceptually, \channel{}s can be viewed as an extension of BLinks~\cite{amir07customizable}, however, unlike BLinks, \channel{}s (1)~do not require messages to be totally ordered at the channel level and (2)~comprise built-in flow control. To forward information, an \channel internally can be divided into multiple subchannels providing first-in-first-out semantics. Each subchannel has a configurable maximum capacity~(i.e.,~an upper bound on the number of messages that can be concurrently in transmission) and relies on a window-based flow-control mechanism to prevent senders from overwhelming receivers. Below, we discuss the specifics of \channel{}s at a conceptual level. For possible implementations please refer to Section~\ref{sec:implementations}.

\begin{figure}
	\includegraphics{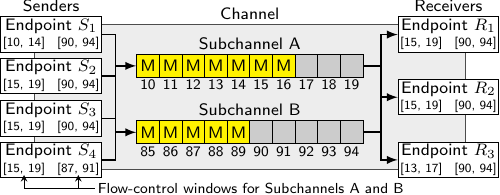}
	\caption{Conceptual view of an example \channel with two independent subchannels that both have a maximum capacity of ten messages~(M). Senders~($S_*$) and receivers~($R_*$) access the subchannels via their local endpoints; each endpoint manages its own subchannel-specific flow-control windows.}
	\label{fig:channel}
\end{figure}

\headline{Overview}
%
Figure~\ref{fig:channel} presents an example \channel that comprises two subchannels and connects four senders to three receivers. Subchannels of the same \channel are independent of each other and can be regarded as distributed queues with limited capacity that distinguish messages based on unique position indices. Both senders and receivers run dedicated endpoints which together form the \channel and enable the replicas to access it. When a replica sends a message, it provides its local endpoint with the information which subchannel and position to use for the message~(\texttt{send()}, see Figure~\ref{fig:interfaces}). Similarly, to receive a message a replica queries its local endpoint for the message corresponding to a specific subchannel and position~(\texttt{receive()}). In addition, \channel endpoints offer a method to shift the flow-control window of a subchannel~(\texttt{move\_window()}), as further discussed below.

\headline{Send Semantics}
%
\channel{}s are not designed to exchange arbitrary messages between replicas but instead provide specific send semantics enabling \system to safely forward the decision of a replica group to another. In particular, tolerating at most $f_s$~senders with Byzantine faults, the \channel only forwards a message after at least $f_s+1$~different senders transmitted a message with identical content using the same subchannel and position. Consequently, in order for a message to pass the channel at least one correct sender must have vouched for the validity of the message's content and requested its transmission. In contrast, messages solely submitted by the up to $f_s$~faulty senders have no possibility of getting through and being delivered to receivers.


\headline{Authentication}
%
\channel{}s protect all channel-internal communication with signatures~(e.g.,~RSA~\cite{rivest78method}, ed25519~\cite{RFC7748}) to enable the recipient of a message to verify the integrity and the origin of the message. If an endpoint is unable to validate the authenticity of a received message, the endpoint immediately discards the message without further processing it.

\headline{Flow Control}
%
With the capacities of subchannels being limited, \channel endpoints apply a flow-control mechanism to coordinate senders and receivers. For this purpose, for each subchannel an endpoint manages a separate window that restricts which messages a sender/receiver is able to transmit/obtain at a given time. If a subchannel's window at a sender endpoint is full, the sender cannot insert additional messages into this subchannel until the endpoint moves the window forward. In the normal case, this action is triggered by receivers calling \texttt{move\_window()} and requesting the start of the window to be shifted to a higher position. Whenever a sender endpoint learns that the window position has changed at one of the receiver endpoints, the sender endpoint sets its own window start to the $f_r+1$~highest position requested by any receiver where $f_r$~denotes the number of receivers with Byzantine faults to tolerate. This ensures that correct sender endpoints only move their windows, and thus discard messages at lower positions, after receiving the information that at least one correct receiver has permitted such a step.

Besides receiver-driven window shifts, our channels also allow senders to request an increase of the starting position of a subchannel's window. If senders opt to do so, it may become impossible for a receiver endpoint to provide the message at the position the endpoint's local replica requested. The same scenario can occur if a receiver endpoint is slow or falls behind~(e.g.,~due to a network problem) while \mbox{$f_r+1$}~other receivers have already requested the window to be moved forward. In such cases, the affected receiver endpoint aborts the \texttt{receive()} call with an exception and thereby enables its local replica to handle the situation. As discussed in Section~\ref{sec:checkpointing}, replicas react to such an exception by obtaining the missed information from other replicas.

\headline{Use in \system}
%
\channel{}s are an essential building block of \system's modular architecture as they enable us to design a geo-replicated BFT system as a composition of loosely coupled replica groups that interact using the same channel abstraction. In particular, \system relies on two different \channel instances to perform all inter-group communication over long-distance links: the \textit{request channel} and the \textit{commit channel}.
The request channel allows an execution group to forward newly received requests to the agreement group; that is, this channel is an \channel that connects $2f_e+1$~senders~(i.e.,~execution replicas) to $3f_a+1$~receivers~(i.e.,~agreement replicas). To transmit the requests, the request channel comprises multiple subchannels, one for each client. In contrast, the commit channel only consists of a single subchannel and is used by the agreement group to inform an execution group about the totally ordered sequence of agreed requests. The commit channel consequently is responsible for forwarding the decisions of $3f_a+1$~senders to $2f_e+1$~receivers. In summary, the agreement group maintains a pair of \channel{}s~(i.e.,~one request channel and one commit channel) to each execution group.

\subsection{Application}
\label{sec:application}

Being a generic architecture, \system is compatible with a variety of applications, and its lightweight service interface makes it straightforward to integrate the \system infrastructure with existing implementations. As shown in Figure~\ref{fig:interfaces}, to replicate a service with \system the application needs to provide an \texttt{execute()} method that is called by an execution replica to process a client request in its local application instance and obtain the corresponding result. As is common for actively replicated systems~\cite{schneider90implementing}, \system requires the application to implement a deterministic state machine; that is, when correct execution replicas invoke \texttt{execute()} in the same state and for the same request, then all their application instances must produce the same result and end up in the same state. For checkpointing purposes, the state of an application instance needs to be retrievable~(\texttt{snapshot()}) and adjustable~(\texttt{apply()}) by the infrastructure~(see Section~\ref{sec:checkpointing}). Furthermore, if supplied with knowledge about whether or not a client request updates the application state~(\texttt{is\_read\_only()}), \system is able to improve response times by offering a fast path for read-only requests~(see Section~\ref{sec:protocol}).

\section{Replication Protocol}
\label{sec:replication}

In this section, we provide details on the replication protocol that runs at the core of the \system architecture. Specifically, we focus on \system's mechanisms for ordering and processing client requests~(see Section~\ref{sec:protocol}), creating checkpoints in the agreement group as well as the execution group~(see Section~\ref{sec:checkpointing}), establishing a global flow control~(see Section~\ref{sec:flow}), dynamically adapting to workload changes~(see Section~\ref{sec:adaptability}), and tolerating faulty clients and replicas~(see Section~\ref{sec:faulty}).

\subsection{Request Handling}
\label{sec:protocol}

\begin{figure}
	\includegraphics{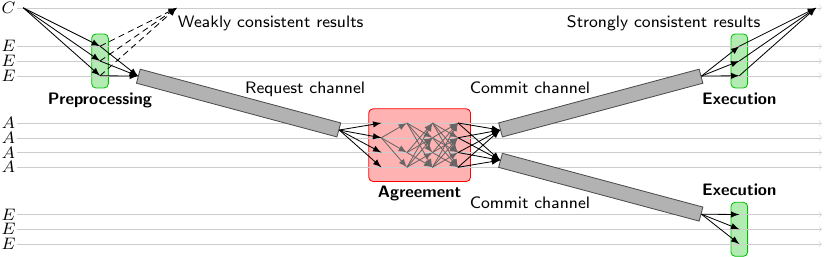}
	\caption{Overview of \system's replication protocol}
	\label{fig:protocol}
\end{figure}

\system differentiates between requests that potentially modify application state~(``writes'') and those that do not~(``reads''). This distinction enables the system to handle requests of each category as efficiently as possible. While writes need to be applied to all execution groups to keep their states consistent, it is sufficient for reads to only process them at the execution group a client is connected to. Figure~\ref{fig:protocol} gives an overview of how requests flow through \system. Below, we provide details on the system's replication protocol for writes and reads. In this context, it is important to note that all messages exchanged between clients and replicas must be authenticated, for example using HMACs~\cite{tsudik92message}. For messages sent through \channel{}s, the authentication is handled by the channels. A formal proof of the protocol's correctness and liveness can be found in~\cite{eischer20resilient-extended}.


{%
\def\pseudocode{\normalfont}
\font\lstttx=rm-lmtl10 scaled 820
\font\lstbttx=rm-lmtk10 scaled 820
\lstset{
	basicstyle=\linespread{.8}\footnotesize\lstttx,
	keywordstyle=\pseudocode\bfseries,
	emphstyle=\lstbttx,
	commentstyle=\commentsize\textit,
	tabsize=2,
	numberstyle=\scriptsize,
	numbersep=2.2mm,
	xleftmargin=5mm,
	numbers=left,
	frame=none,
	columns=fullflexible,
	numberblanklines=true,
	emptylines=2,
	breaklines=true,
	breakatwhitespace=false,
	escapechar=\%,
	mathescape,
	morecomment=[l]{//},
	morecomment=[s]{/*}{*/},
	morestring=[b]",
	aboveskip=-3mm,
	morekeywords={if,else,while,for,each,in,parallel,sleep,until,return,interface,on,callback,event,main,loop,true,false},
	emph={send, receive, move_window, execute, snapshot, apply, order, ordered, collect_garbage},
	emphstyle=[2]\rmfamily\scshape,
	moreemph=[2]{Write,Request,TooOld,Result,SeqNr,Execute},
}%
\begin{figure}
	\vspace{2mm}\hrule\pseudocode\commentsize{\begin{center}Execution Replica of Execution Group~$\mathcal{E}$\end{center}}\hrule\vspace*{2mm}\vspace{8pt}
	\begin{lstlisting}
$s_n$ := $0$%\hfill%// Current sequence number
$t[c]$ := $0$%\hfill%// Vector with counter value of latest forwarded client request
$u[c]$ := $\varnothing$%\hfill%// Result cache $\langle \textsc{Result}, u_c, t_c\rangle$%\vspace{2pt}\hrule{}\vspace{3pt}%
on receive($r$ = $\langle\textsc{Write},w,c,t_c\rangle$ %\pseudocode{from}% $c$):
	if $t_c \leq t[c]$: return%\pseudocode{ send result $u[c]$ to $c$}%%\label{code:exec-cache-start}%
	$t[c]$ := $t_c$%\hfill%// Remember forwarded request%\label{code:exec-cache-end}%
	request-IRMC.move_window($c$, $t_c$)%\label{code:exec-send-start}%
	request-IRMC.send($c$, $t_c$, $\langle \textsc{Request},r,\mathcal{E}\rangle$)%\label{code:exec-send-end}\vspace{5pt}%
main loop:
	$m$ := commit-IRMC.receive($0$, $s_n+1$)%\label{code:exec-receive}%
	if $m = \langle \textsc{TooOld}, s'\rangle$: %\pseudocode{fetch checkpoint for $s'$}%
	else:
		$m$ = $\langle\textsc{Execute}, \langle\textsc{Request},\langle\textsc{Write}, w, c, t_c\rangle,\mathcal{E}'\rangle, s_n+1\rangle$
		$s_n$ := $s_n + 1$
		if $(u[c]=\langle\textsc{Result}, *, t_c'\rangle \wedge t_c > t_c') \vee u[c]=\varnothing$:%\label{code:exec-skip}\hfill%// Only execute new requests
			$u_c$ := application.execute($m$)%\label{code:exec-exec-start}%
			%\pseudocode{send $\langle\textsc{Result}, u_c, t_c\rangle$ to $c$}% if $\mathcal{E}' = \mathcal{E}$%\pseudocode{ and store in $u[c]$}%%\label{code:exec-exec-end}%
		if $s_n \equiv 0~mod~k_e$:%\label{code:exec-cp-gen-start}%
			%\pseudocode{create checkpoint for $s_n$ with $u$ and }%application.snapshot()%\vspace{5pt}%
on %\pseudocode{stable checkpoint}%(SeqNr $s$, $u'$, $st$):%\label{code:exec-cp-start}%
	commit-IRMC.move_window($0$, $s+1$)
	if $s \geq s_n$: $s_n$ := $s$; application.apply($st$); $u$ := $u'$%\label{code:exec-cp-end}\vspace{3mm}\hrule\pseudocode{\begin{center}Agreement Replica\end{center}}\hrule\vspace*{2mm}%
$s_n$ := $0$%\hfill%// Current sequence number
$t[c]$ := $0$%\hfill%// Counter values of latest agreed request; used by consensus
$t^+[c]$ := $0$%\hfill%// Counter values for next expected request
$\textrm{\lstttx{AG-WIN}} \geq k_a$%\hfill%// Size of agreement window
win := [1,AG-WIN]%\hfill%// Range with [lower, upper] bound (inclusive)
hist := %\pseudocode{last $|${\lstttx{commit-IRMC}} window$|$ \textsc{Execute} messages}%%\vspace{2pt}\hrule{}\vspace{3pt}%
for each %\pseudocode{client $c$ and execution group $\mathcal{E}$}% in parallel:
	while true:
		$m$ := request-IRMC.receive($c$, $t^+[c]$)%\pseudocode{ from group }%$\mathcal{E}$%\label{code:ag-receive}%
		if $m$ = $\langle\textsc{TooOld}, t_c\rangle$: $t^+[c]$ := $t_c$%\label{code:ag-receive-tooold}%
		else: // $m$ = $\langle\textsc{Request},\langle\textsc{Write}, w, c, t_c\rangle,\mathcal{E}\rangle$
			agreement.order($c$, $t_c$, $m$)%\label{code:ag-order}%
			$t^+[c]$ := $t_c + 1$%\vspace{5pt}%
on agreement.ordered(SeqNr $s$, $r$ = $\langle\textsc{Request},\langle\textsc{Write},w,c,t_c\rangle,\mathcal{E}\rangle$):%\label{code:ag-deliver}%
	sleep until%\pseudocode{ upper limit of {\lstttx{win}} $\geq s$}%%\label{code:ag-win-sleep}%
	$s_n$ := $s$
	$t[c]$ := $t_c$
	$t^+[c]$ := $\max(t_c + 1, t^+[c])$
	commit-IRMC.send($0$, $s$, $\langle\textsc{Execute}, r, s\rangle$) for each%\pseudocode{ execution group $\mathcal{E}$ and add }%Execute%\pseudocode{ to \lstttx{hist}}%%\label{code:ag-execute}%
	if $s_n \equiv 0~mod~k_a$:%\label{code:ag-cp-gen-start}%
		%\pseudocode{create checkpoint for $s_n$ with $t$, \lstttx{hist}}%%\label{code:ag-cp-gen-end}\vspace{5pt}%
on %\pseudocode{stable checkpoint}%(SeqNr $s$, $t'$, hist'):
	commit-IRMC.move_window($0$, $s-|\textrm{\lstttx{hist'}}|+1$) for each%\pseudocode{ execution group $\mathcal{E}$}%
	agreement.collect_garbage($s+1$, $t'$)%\label{code:ag-cp-gc}%
	if $s > s_n$:
		h_missing := $\{\langle\textsc{Execute}, r, s'\rangle \in$ hist'$\,|\,s' \in [s_n+1, s]\}$
		%\pseudocode{apply checkpoint to $s_n$, $t$ and \lstttx{hist}}%%\label{code:ag-cp-apply-start}%
		for each%\pseudocode{ execution group $\mathcal{E}$}%:
			%\pseudocode{send {\lstttx{h\_missing}} via {\lstttx{commit-IRMC}} of group $\mathcal{E}$}%%\label{code:ag-cp-apply-end}%
	win := [$s+1$, $s+\textrm{\lstttx{AG-WIN}}$]%\label{code:ag-win-move}\label{code:ag-cp-end}%
	\end{lstlisting}
	\hrule
	\caption{\system protocol for writes (pseudo code)}
	\label{def:pseudo-write}
\end{figure}
}%

\headline{Writes}
%
\system's protocol for writes is presented in Figure~\ref{def:pseudo-write} as pseudo code.
To perform a write operation~$w$, a client~$c$ creates a corresponding message \msg{Write}{w, c, t_c} using a unique client-local counter value~$t_c$ and sends the message to all replicas of an execution group. In general, a client for this purpose may select any execution group in the system, however, in an effort to minimize latency, \system clients typically choose the group closest to their own site.

When an execution replica receives the client's request, it first checks whether the message is correctly authenticated and whether the client has permission to access the system. If any of these checks fail the replica discards the message. Otherwise, the replica of execution group~$\mathcal{E}$ wraps the entire request~$r$ in a message \msg{Request}{r, e} and submits the message to the agreement group via its request channel. More precisely, unless the execution replica  has already forwarded the request~(Lines~\ref{code:exec-cache-start}--\ref{code:exec-cache-end}) it moves the window of the client's subchannel to position~$t_c$ and inserts the write request at that position~(L.~\ref{code:exec-send-start}--\ref{code:exec-send-end}). Once at least $f_e+1$~members of the execution group~(i.e.,~at least one correct execution replica) have validated and forwarded the request, the request channel permits agreement replicas to retrieve the message~(L.~\ref{code:ag-receive}). This allows the agreement group to initiate the consensus process for the message~(L.~\ref{code:ag-order}), which is then performed entirely within the group's region. Having learned that the request is committed and has been assigned the agreement-sequence number~$s$~(L.~\ref{code:ag-deliver}), an agreement replica creates a confirmation \msg{Execute}{r, s}. As write operations need to be processed by all execution groups, the agreement replica sends this message through all commit channels at position~$s$~(L.~\ref{code:ag-execute}).

Once $f_a+1$~agreement replicas~(among them at least one correct replica) have sent an \textsc{Execute} message with the same content and sequence number, a commit channel enables its receivers to obtain the message~(L.~\ref{code:exec-receive}). Having done so, an execution replica processes the included request by applying the corresponding write to its local state~(L.~\ref{code:exec-exec-start}). Each replica of execution group~$\mathcal{E}$ also returns a reply \msg{Result}{u_c, t_c} with the operation's result~$u_c$ to the client that submitted the request with counter value~$t_c$~(L.~\ref{code:exec-exec-end}). The client accepts a result after it has received $f_e+1$~replies with matching result and counter value from different execution replicas.

As we detail in Section~\ref{sec:checkpointing}, when processing writes replicas in \system also create periodic checkpoints~(L.~\ref{code:exec-cp-gen-start}--\ref{code:exec-cp-end} and \ref{code:ag-cp-gen-start}--\ref{code:ag-cp-end}) to assist other replicas that might have fallen behind.

\headline{Reads}
%
For reads, \system offers two different operations providing weakly consistent and strongly consistent results, respectively. To perform a weakly consistent read, a client sends a read request to all members of an execution group, which for a valid request immediately responds with a result, as illustrated by the dashed lines in Figure~\ref{fig:protocol}. As for writes, a client verifies the result based on $f_e+1$ matching replies. Weakly consistent reads achieve low latency as they only involve communication between the client and its execution group. Due to these reads being processed without further coordination with writes, in the presence of concurrent writes to the same state parts they may return stale values or fewer than $f_e+1$ matching results, similar to the optimized reads in existing BFT protocols~\cite{castro99practical,sousa15separating}. \system clients react to stalled reads by retrying the operation or performing a strongly consistent read, which is guaranteed to produce a stable result.

Strongly consistent reads in \system for the most part have the same control and data flow as writes, with one important exception. With reads not modifying the application state, it is sufficient to process them at the client's execution group. Consequently, after a read request completed the consensus process, agreement replicas only forward it to the execution group that needs to handle the request. The \textsc{Execute}s to all other groups instead contain a placeholder including only the client request counter value for the same sequence number, thereby minimizing network and execution overhead.

\headline{Extensions for Sequential Consistency}
%
Apart from built-in support for trading off consistency for efficiency, the \system architecture also offers the flexibility to efficiently implement other consistency guarantees for reads. To illustrate this aspect, in the following we show how \system can be extended to provide sequential consistency. By definition, this means that all operations must appear to happen according to a single total order, such that the operations of an \emph{individual} client are processed in the order they were issued by the client~\cite{attiya94sequential, viotti16consistency}. However, it is for example possible for one client to complete a write and for another client to still be able to read an older state afterwards. In contrast to strong consistency~(i.e.,~linearizability~\cite{herlihy90linearizability}), sequential consistency only requires the order of operations for each individual client to match the order in which the requests were issued, but does not require that order to hold across operations from different clients.

As read requests by definition have no side effects and therefore are only relevant for the requesting client, it is sufficient for the total order to only include write requests. The only restriction from a client's point of view is that after reading or writing a value it must always see the same or a later state. With writes in \system already being totally ordered, we thus only have to guarantee that for a client weakly consistent reads are processed after all earlier write requests of the same client (\emph{read-my-writes}) and that a read accesses the same or a newer state than all of the client's previous reads (\emph{monotonic reads})~\cite{viotti16consistency}.

In order to solve the read-my-writes requirement, similar to \textsc{Weave}~\cite{eischer20low-latency}, a client adds a minimum sequence number~$s_{min}$ to its weakly consistent read request, which specifies that a replica's execution must have advanced to at least $s_{min}$ before performing the read. In addition, the reply to a write request is extended with the sequence number~$s_{write}$ at which the write took effect. A client now has to wait for $f_e+1$~replies to its write containing the same result and sequence number. As all correct replicas execute an operation at the same sequence number, both their results and sequence numbers are identical, which guarantees that a client will be able to receive $f_e+1$~matching replies. After accepting a result for a write, the client updates its minimum sequence number~$s_{min}$ to $s_{write}$. This guarantees that correct replicas will only reply to future reads once they have executed the write, and it also ensures that the client does not accept older replies from faulty replicas.

To satisfy the monotonic-reads requirement, a client's subsequent requests need to include the sequence number of the latest system state the client has observed in earlier operations. However, as weakly consistent read requests are not ordered, there is no specific sequence number at which such a request will be executed. Fortunately, as a substitute replicas can respond with the sequence number~$s_{update}$ of the latest state change that had an influence on the read's result. In case of a key-value store application, this for example could be the sequence number at which the value for a key was last updated. Similar to write requests, a client waits for $f_e+1$~matching replies with identical results and sequence numbers. As~$s_{update}$ is likely to change much less frequently than the latest sequence number a replica has processed, there is an increased chance that all correct replicas return the same sequence number and that consequently the necessary quorum of matching replies will be reached. Once this is the case, the client only updates its minimum sequence number~$s_{min}$ if the received sequence number~$s_{update}$ is larger. On the other hand, if a client does not accept a reply within a timeout, the client issues the request again as a strongly consistent read.
 
\subsection{Checkpointing}
\label{sec:checkpointing}

As discussed in Section~\ref{sec:channels}, an \channel{} may garbage-collect messages before they have been delivered to all correct receivers. In the normal case in which all receivers advance at similar speed, this property usually does not take effect, resulting in each receiver to obtain every message. To address exceptional cases in which a correct receiver misses messages~(e.g.,~due to a network problem), \system provides means to bring the affected receiver up to date via a checkpoint. The specific contents of a checkpoint vary depending on the receiver-replica group~(see below).
Checkpoints are periodically created after a group has agreed on\,/\,processed the message for a sequence number~$s$ that satisfies $s \equiv 0~mod~k$. The checkpoint interval~$k$ of a replica group is configurable and for the execution to sustain liveness must be smaller than the maximum capacity of the group's input \channel. The agreement-checkpoint interval~$k_a$ may be selected independently from the interval for execution checkpoints~$k_e$.

\headline{Agreement Checkpoints}
%
Having completed the consensus process for a request for which a checkpoint is due, an agreement replica creates an agreement snapshot and includes (1)~a vector~$t$ that for each client contains the counter value~$t_c$ of the client's latest agreed request and (2)~the last \textsc{Execute} messages corresponding to the commit subchannel capacity~(L.~\ref{code:ag-cp-gen-start}--\ref{code:ag-cp-gen-end} in Figure~\ref{def:pseudo-write}). In a next step, the agreement replica computes a hash~$h$ over the snapshot and sends a message \msg{Checkpoint}{h, s} which is protected with a signature to all members of its group. Having obtained $f_a+1$~correctly signed and matching checkpoint messages for the same sequence number, a replica has proof that its snapshot is correct. At this point, the replica can move forward its separate window used to ensure the periodic creation of a new checkpoint~(L.~\ref{code:ag-win-sleep} and \ref{code:ag-win-move}) and also instruct the consensus protocol to garbage-collect preceding consensus instances~(L.~\ref{code:ag-cp-gc}).

Agreement replicas require periodic checkpoints to continue ordering new requests and thus there is at least one correct agreement replica that possesses both a corresponding valid checkpoint as well as proof of the checkpoint's correctness in the form of $f_a+1$~matching checkpoint messages.
As a consequence, if a correct agreement replica falls behind and queries its group members for the latest checkpoint, the replica will eventually be able to acquire this checkpoint, verify it, and apply it in order to catch up by skipping consensus instances. In such case, the checkpoint enables the replica to learn (1)~the request-subchannel positions at which to query the \channel for the next client requests and (2)~the \textsc{Execute}s of the skipped consensus instances~(L.~\ref{code:ag-cp-apply-start}--\ref{code:ag-cp-apply-end}).

\headline{Execution Checkpoints}
%
Execution-group checkpointing follows the same basic work flow as in the agreement group. An execution snapshot comprises a copy of the application state and the latest reply to each client, similar to the checkpoints in Omada~\cite{eischer19scalable}. This information enables a trailing execution replica to consistently update its local state without needing to process all agreed requests. When an execution checkpoint for a sequence number~$s$ becomes stable at an execution replica, the replica moves the flow-control window of its incoming commit channel to $s+1$~(L.~\ref{code:exec-cp-start}--\ref{code:exec-cp-end}).
This ensures that agreed requests are only discarded after at least one correct execution replica has collected a stable checkpoint.
Note that there is no need for checkpoints to contain requests.
A client moves its request subchannel's window forward by issuing a new request, thereby confirming that the old request can be garbage-collected from the \channel.
This also allows execution replicas to skip forward to the current request~(L.~\ref{code:ag-receive-tooold}).

\subsection{Global Flow Control}
\label{sec:flow}

With the flow-control mechanism of an \channel only operating at the communication level between two replica groups, \system takes additional measures to coordinate the message flow at the point where the endpoints of multiple \channel{}s meet: the agreement group. Specifically, there are two types of messages~(i.e.,~new requests received through request channels and \textsc{Execute}s sent through commit channels) that have individual characteristics and are handled in different ways.

\headline{Inputs}
%
With regard to incoming requests, agreement replicas represent the receiver side of request channels and therefore directly manage the positions of the channels' flow-control windows. As described in Section~\ref{sec:checkpointing}, to be able to quickly retrieve new requests an agreement replica updates the counter value of each client's latest request each time an agreement checkpoint becomes stable.

\headline{Outputs}
%
With regard to outgoing \textsc{Execute}s, in contrast, agreement replicas represent the sender side of commit channels and therefore depend on the respective execution group at the other end of each channel to move the flow-control window forward. To prevent a single execution group from delaying overall progress, agreement replicas in \system do not wait until they are able to submit a newly produced \textsc{Execute} to every outgoing commit channel. Instead, having completed inserting an \textsc{Execute} for a sequence number~$s$ into $n_e-z$~commit channels an agreement replica is allowed to continue; $n_e$~is the total number of execution groups in the system and $z$ a configurable value~($0 \leq z < n_e$). To inform the execution groups of trailing commit channels, once such a request is garbage-collected a replica updates the channels' window positions to sequence number~\mbox{$s+1$}. If an affected execution replica subsequently tries to receive \textsc{Execute}s for sequence numbers of $s$ or lower, the commit channel responds with an exception~(see~Section~\ref{sec:channels}). In reaction, the execution replica starts to seek a stable execution checkpoint, querying members of both its own group and others, in order to compensate for the missed messages.

\subsection{Adaptability}
\label{sec:adaptability}

\system's modular architecture makes it possible to dynamically change the number of execution groups in the system and thereby adjust to varying workloads. With the consensus protocol being limited to the agreement group, in contrast to traditional BFT systems such a reconfiguration in \system does not require complex mechanisms or subprotocols.

\headline{Adding an Execution Group}
%
To add a new execution group~$\mathcal{E}$ to the system, a privileged admin client first starts the replicas of the group and then submits an \msg{AddGroup}{\mathcal{E}, \mathcal{I}} message; $\mathcal{I}$ is a set containing the identity and address of each group member. As soon as the agreement process for this message is complete, agreement replicas establish an \channel{} pair~(i.e.,~a request channel and a commit channel) to the new execution group, update the execution-replica registry to reflect the changes, and start the reception of requests and the forwarding of \textsc{Execute}s. Trying to obtain an \textsc{Execute} for sequence number 0, the new replicas will be notified by their commit channels that they have fallen behind and consequently use the mechanism of Section~\ref{sec:flow} to fetch an execution checkpoint from another group. 
\pagebreak

\headline{Removing an Execution Group}
%
To remove an existing execution group~$\mathcal{E}$ from the system, the administrator submits a \msg{RemoveGroup}{\mathcal{E}} message that, once agreed on, causes the agreement replicas to update the execution-replica registry and close their \channel{}s to the affected group.
The agreement replicas only accept the removal message if $n_e - z \geq 2$, that is at least two execution groups must store the current state.
This ensures that after removing an execution group at least one execution group with the current state will remain.

\subsection{Handling Faulty Clients and Replicas}
\label{sec:faulty}

Besides enabling \system's modular architecture, \channel{}s also play a crucial role when it comes to limiting the impact faulty clients and replicas can have on the system. In this context, especially one \channel property is of major importance: the fact that a channel only delivers a message after $f+1$~senders submitted it and the channel therefore has proof that at least one correct sender vouches for the message's validity~(see Section~\ref{sec:channels}). If, for example, a faulty client either sends conflicting requests to an execution group or the same request to fewer than \mbox{$f_e+1$}~execution replicas, the request channel of the affected execution group prevents the message's delivery to the agreement group. Note that in such case the effects of the faulty client are strictly limited to the subchannel of this client, which will not deliver a request if fewer than $f_e+1$~execution replicas insert the same message. As execution replicas use a dedicated request subchannel for each client, the subchannels of correct clients remain unaffected.

If faulty execution replicas collaborate with a faulty client, different agreement replicas may receive different values for this client's requests.
For example, a faulty client might submit a different request~$R_1$, $R_2$, \dots, $R_{f_e+1}$ to each of the $f_e+1$~correct execution replicas of one group and provide all requests to the $f_e$~faulty execution replicas of that group.
Depending on which of the request versions the faulty execution replicas transmit to which agreement replica, in such a situation it is possible that some agreement replicas obtain an $f_e+1$~quorum for request~$R_1$ while others receive $f_e+1$~matching messages for request~$R_2$ and so on.
Again, the effects are limited to the faulty client's subchannel; requests of correct clients can proceed as usual.
This scenario is not specific to \system, but in a similar way can also occur in traditional BFT systems~\cite{castro99practical,yin03separating,veronese10ebawa,sousa15separating,eischer19scalable}, in which clients directly submit their possibly conflicting requests to the replicas performing the agreement.
Consequently, all BFT protocols that tolerate faulty clients already comprise mechanisms to handle this scenario.
Execution replicas in \system additionally only execute client requests with a counter value which is higher than the highest value processed so far for that client, which ensures that old or duplicate requests are skipped.

Besides tolerating faulty clients, agreement protocols in general also provide means that allow correct follower replicas to elect a new leader if the current leader is faulty and, for example, fails to start the consensus process for a new client request within a given timeout~\cite{castro99practical,yin03separating,veronese10ebawa,sousa15separating,eischer19scalable}. To be able to monitor the leader, follower replicas must obtain information about incoming requests. In \system, this is ensured by the fact that request channels only garbage-collect a request from a correct client if the latter has successfully obtained a valid reply. A request for which this is not the case will be uploaded to all correct members of the client's execution group and through this group's request channel eventually reach all correct follower agreement replica, thereby enabling followers to hold the leader accountable.

Finally, faulty agreement replicas cannot forward manipulated messages via the commit channel.
As the consensus process ensures that all correct agreement replicas deliver the same total order of requests, eventually $f_a+1$~correct agreement replicas will send matching messages  enabling the execution groups to receive the correctly ordered requests.
In contrast, the delivery of faulty requests sent by the faulty agreement replicas is prevented by the \channel.

\section{\channel Implementations}
\label{sec:implementations}

In this section, we present two different variants to implement inter-regional message channels, focusing on simplicity~(\channela) and efficiency~(\channelb), respectively. Additional variants are possible, as discussed in Section~\ref{sec:related}.

\vspace{2mm}

\begin{figure}[b]
	\subfloat[\channela\label{fig:implementations-a}] {
		\includegraphics[page=1]{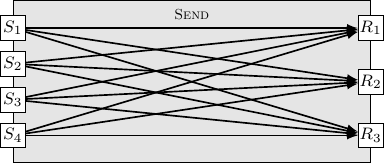}
	}
	\hspace{2mm}
 	\subfloat[\channelb\label{fig:implementations-b}] {
		\includegraphics[page=2]{implementations.pdf}
	}
	\caption{Overview of two possible \channel implementations.}
	\label{fig:implementations}
\end{figure}

\headline{\channel with Receiver-side Collection~(\channela)}
%
The receiver endpoint of an \channel only delivers a message~$m$ for a specific subchannel~$sc$ and position~$p$ if at least $f_s+1$~senders previously instructed the channel to transmit a message with identical content for the same subchannel position~(see Section~\ref{sec:channels}). As illustrated in Figure~\ref{fig:implementations-a}, the \channela solves this problem by each sender endpoint~$S_x$ directly forwarding a \smsg{Send}{m, sc, p}{S_x, \mathcal{X}} message and thereby enabling each receiver endpoint to individually collect $f_s+1$ matching messages. To allow receivers to verify the origin and integrity of a \textsc{Send}, a sender signs messages with its private key~$\mathcal{X}$.
When a receiver requests a subchannel's flow-control window to be shifted, its receiver endpoint~$R_y$ submits a signed \smsg{Move}{sc, p}{R_y, \mathcal{Y}} message to all sender endpoints. For each receiver and subchannel, a sender endpoint stores the \textsc{Move} message with the highest position~$p$ and sets the subchannel's window start to the $f_r+1$~highest position requested by any receiver~(see Section~\ref{sec:channels}).
To request a shift of a subchannel's flow-control window, sender endpoints also send \textsc{Move} messages which the receivers process analogously.


\vspace{2mm}

\headline{\channel with Sender-side Collection~(\channelb)}
%
\channelb{}s minimize the number of messages transferred across wide-area links by applying the concept of \emph{collectors}~\cite{gueta19sbft}. That is, sender endpoints in \channelb{}s do not submit their \textsc{Send}s to the receiver side but, as indicated in Figure~\ref{fig:implementations-b}, instead exchange signed hashes of them within the sender group. Each sender endpoint serves as a collector, which means that the endpoint assembles a vector~$\vec{v}$ of $f_s+1$~correct signatures from different senders for the same \textsc{Send} message content~$sm$. Having obtained this vector, a collector~$S_x$ sends it in a signed \smsg{Certificate}{sm, \vec{v}}{S_x, \mathcal{X}} message to one or more receiver endpoints. On reception, a receiver verifies the validity of the \textsc{Certificate} by checking both the signatures of the message and the $f_s+1$~signatures contained in the vector~$\vec{v}$. If all of these signatures are correct and match the \textsc{Send} message content~$sm$, the endpoint has proof that $sm$ is valid as it was sent by at least one correct replica and delivers the associated message to its receiver on request.

\channelb receiver endpoints individually select the sender endpoint serving as their current collector and announce these decisions attached to their \textsc{Move}s. As a protection against faulty collectors, all sender endpoints periodically transmit \smsg{Progress}{\vec{p}}{S_x, \mathcal{X}} messages directly to receiver endpoints in which they include a vector~$\vec{p}$ with the highest position of each subchannel for which they have a \textsc{Certificate}. If at least $f_s+1$~sender endpoints claim to have reached a certain position but a receiver's collector fails to provide a corresponding and valid \textsc{Certificate} within a configurable amount of time, the endpoint switches to a different collector.

\section{Optimizations}
\label{sec:optimizations}

In the following, we present a number of optimizations that enable \system to provide its service both more effectively as well as efficiently. With the authentication of messages being one of the most computationally expensive responsibilities of a replica, several of these optimizations focus on reducing the overhead for signature creation or verification. 


\begin{figure}[b]
	\includegraphics{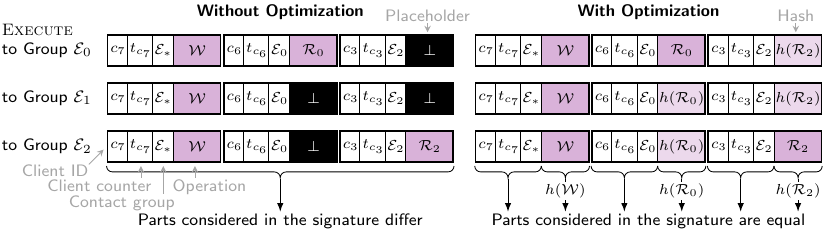}
	\caption{Impact of the signature-sharing optimization on \textsc{Execute} messages to different execution groups~$\mathcal{E}_*$ by example of a committed request batch that consists of a write~$\mathcal{W}$ issued by a Client~$7$, a strongly consistent read~$\mathcal{R}_0$ issued by Client~$6$ to $\mathcal{E}_0$, and another strongly consistent read~$\mathcal{R}_2$ issued by Client~$3$ to $\mathcal{E}_2$.}
	\label{fig:signature-sharing}
\end{figure}

\subsection{Signature Sharing between IRMCs}
\label{sec:signature-sharing}

The first optimization aims at reducing the number of signatures an agreement replica has to compute in order to send authenticated \textsc{Execute}s through its commit channels. Specifically, the optimization addresses scenarios in which a combination of two aspects otherwise would lead to inefficiency: (1)~the fact that committed requests for strongly consistent reads should only be transmitted to the group that the corresponding client is connected to (see Section~\ref{sec:protocol}) and (2)~agreement protocols typically reaching consensus on request batches instead of single requests~\cite{castro99practical}.

As shown in Figure~\ref{fig:signature-sharing}, without further measures applying these two techniques would result in agreement replicas having to generate different signatures for different groups due to the contents of \textsc{Execute} messages differing. The example in the figure shows a scenario in which three requests have been batched together: a write~$\mathcal{W}$ that needs to be processed by all execution groups, a read~$\mathcal{R}_0$ that is solely intended for an execution group~$\mathcal{E}_0$, and a read~$\mathcal{R}_2$ to be performed at an execution group~$\mathcal{E}_2$. With full read requests only being sent to their respective groups and otherwise substituted with placeholders, each execution group in such a situation should be supplied with a dedicated \textsc{Execute}. Consequently, authenticating messages in the traditional way~(i.e.,~by considering the entire message contents when computing signatures) would lead to each \textsc{Execute} requiring a different signature, thereby increasing authentication overhead.

To circumvent this problem, our optimization constructs \textsc{Execute}s in such a way that it is possible for an agreement replica to compute a signature that can be shared across all the replica's commit channels. As shown on the right side of Figure~\ref{fig:signature-sharing}, for this purpose we use a hash~$h(\mathcal{R}_*)$ of the original operation~$\mathcal{R}_*$ instead of a generic placeholder. Furthermore, for all operations in a batch we consider their hashes when computing the \textsc{Execute}'s signature, not the full operations. As a key benefit, this approach results in identical signatures for all of an agreement replica's \textsc{Execute}s and therefore allows a replica to save resources by performing the signature computation only once per \textsc{Execute}, independent of the number of execution groups the replica is connected to.

\subsection{Signature Batching}
\label{sec:signature-batching}

\vspace{-.7mm}

To further reduce the cost for signature computations, our second optimization combines multiple messages before sending them through an \channel, thereby enabling a replica to only generate a single signature for the entire batch instead of one signature for each message in it. That is, in contrast to the first optimization presented in Section~\ref{sec:signature-sharing}, which leverages similarities across multiple \channel{}s, our second optimization targets individual \channel{}s. Consequently, it is applicable to both request channels as well as commit channels. With \textsc{Execute}s typically including request batches~(see Section~\ref{sec:signature-sharing}), in general there already is an implicit batching effect for commit channels, which is why in the following we focus on describing our approach in the context of request channels.

Request channels pose a special challenge when it comes to batching as they are used at a point in the protocol at which there does not yet exist a total order on the transmitted requests. In particular, with client requests commonly arriving at execution replicas at different times and in different orders, there is no easy way to ensure that all correct execution replicas batch the same requests together. To account for this fact and still be able to individually verify specific requests, we apply a technique similar to the one used to sign unordered responses in Basil~\cite{suri-payer21basil}. It is based on the idea of each sender endpoint grouping its messages in a Merkle tree~\cite{merkle79secrecy} and only signing the tree root to amortize the signature costs over all messages in the tree. Following this approach, the tree provides proof that a certain message is part of the tree signed by the signature, which in turn indirectly proves the validity of each message contained in the tree.

Applying the batching optimization, \channel{}s in \system use Merkle trees as follows: A sender endpoint collects multiple messages to be sent on any of the \channel{}'s subchannels, assembles them into a Merkle tree, and then signs the root hash. Next, the sender attaches $\langle mc, i, \vec{h}, sig \rangle$ as signature to each message, where $mc$ is the number of messages included in the tree, $i$ is the message index representing the message's position in the tree, $\vec{h}$ is the list of intermediate hashes necessary to verify the path from the message to the tree root, and $sig$ is the signature of the root hash.
To ensure that the combination of the message index~$i$ and the message count~$mc$ uniquely defines the path from a message to the tree root, we define the structure of the tree to only depend on the message count. Specifically, the messages are included in a complete binary tree, that is all levels except the last are completely filled and the last level is filled from left to right. Having constructed the tree based on these rules, a sender then assigns the messages to leaf nodes. After the reception of a message, the receiver endpoint has to verify that the message is part of the Merkle tree signed by the signature and that the signature is valid. The result of the signature check should be cached to avoid multiple verifications of the signature when different messages belonging to the same Merkle tree arrive. The size of the signature-check cache for each sender replica should be limited to the amount of slots an \channel can store at a time.

\vspace{-.7mm}

\subsection{Client Request Verification Offloading}

\vspace{-.7mm}

To ensure that the output of the agreement group solely consists of valid client requests, agreement replicas must only start the consensus process for a request if they have confirmation that the request is correctly authenticated. A straightforward way to fulfill this requirement is to force agreement replicas to verify the signatures of all incoming client requests themselves, however such an approach can be associated with significant computation overhead. This is especially true if one or both of the following aspects apply: (1)~If throughput is high, the signatures of several thousands of requests per second might have to be verified. (2)~Depending on the authentication scheme in use, the verification of a single request already can be comparably expensive. In our test environment, for example, the duration of a verification procedure increases from 14~$\mu{}s$ to 173~$\mu{}s$ when relying on ed25519~\cite{RFC7748}, a modern elliptic-curve-based signature, instead of RSA~\cite{rivest78method}. This means that for ed25519 verification takes even longer than signature creation~(80~$\mu{}s$ on average).

To eliminate this potential computational bottleneck in the agreement group, we exploit the fact that execution replicas already check the validity of incoming client requests prior to forwarding them via the request channel. Instead of verifying the request signature at the agreement group, it is therefore sufficient to verify that a request was transferred by the \channel, which guarantees that at least one correct execution replica has successfully verified the request. For this purpose, we enable agreement replicas to assemble a \emph{proof of transfer} that certifies the transmission of a message through the request channel. Requests for which no such proof exists~(e.g.,~when a faulty agreement leader tries to forge client requests) are not considered for consensus by correct agreement replicas.

Agreement replicas construct a proof of transfer as part of an additional protocol phase that is executed prior to the consensus process and only involves group-internal communication. As a first step, whenever an agreement replica  obtains a request~$q$ via the request channel, the replica creates a request descriptor $d = \langle c, t_c, h(q) \rangle$ consisting of the client id~$c$, the client-counter value~$t_c$, and a hash~$h(q)$ of the request's content. For a batch $\vec{d}$ of multiple such descriptors, the replica then broadcasts a MAC-authenticated $\langle \textsc{Verify}, \vec{d}\rangle$ message to all other agreement replicas. If a replica this way learns a request descriptor that the replica itself so far has not yet obtained, but which is part of at least $f_a+1$ \textsc{Verify}s sent by other replicas, the replica includes the descriptor in its own next \textsc{Verify}. The proof of transfer for a request is complete once a replica has received $2f_a+1$ \textsc{Verify}s from different replicas containing the same descriptor for the request. At this point, each replica that obtained the request via a request channel then passes it on to the agreement protocol.

\subsection{Improved Availability for Weakly Consistent Reads}

As discussed in Section~\ref{sec:protocol}, a weakly consistent read completes at a client's local execution group as long as $f_e + 1$~execution replicas have equal application states when processing the corresponding request and therefore return matching results.
If a client does not receive sufficient matching replies, it can fall back to submitting a strongly consistent read which is guaranteed to complete.
Without the need for coordination with other requests, weakly consistent reads usually also succeed in the unlikely event that an execution group is disconnected from the agreement group. An interesting corner case in this context are scenarios in which a network partition between the agreement group and an execution group affects execution replicas gradually, and therefore leads to some execution replicas making less progress than others with regard to the most recent writes. In general, this does not pose a major problem because (1)~the deviations between execution replicas are temporary and automatically resolved once the execution group is reconnected to the agreement group, (2)~the effect is limited to the typically small number of state parts at which the latest changes occurred, and (3)~a client always has the option to resort to a different execution group in case a weakly consistent read does not complete at its local group.

Alternatively, the issue can be avoided altogether by ensuring that eventually all correct execution replicas of a client's local group will execute the same set of writes even if the wide-area communication to the agreement group is interrupted. To do so, we extend the commit channels' receiver endpoints with a mechanism that allows them to regularly exchange status messages about their latest progress; for each subchannel, progress is represented by the highest position up to which the receiver has collected valid certificates for this and all earlier sequence numbers. Based on these status messages, a receiver is able to detect whether another receiver has fallen behind and (if necessary) forward the certificates that the other receiver is missing. Due to the certificates being signed, the trailing receiver is able to verify their validity before accepting them. To avoid duplicate message transfers, an endpoint waits for a short timeout before forwarding a certificate. If the other endpoint has received the certificate in the meantime and reports so in a status message, then no forwarding takes place. In summary, the optimization enables an execution replica to eventually learn certificates for all sequence numbers delivered by any other correct execution replica in the group, thereby ensuring that at some point all correct replicas will provide the same results to weakly consistent reads even if the wide-area communication between agreement group and execution group is interrupted for a long period of time.

\section{Evaluation}

In this section, we experimentally evaluate \system in comparison to existing approaches for BFT wide-area replication. Furthermore, we conduct an extensive analysis of our two \channel variants.
 
\subsection{Environment}
%
To compare different techniques, we implemented a Java-based prototype that can be configured to reflect three different system architectures~(cf.~Section~\ref{sec:background}): (1)~\textbf{\bft} represents the traditional approach of distributing a single set of replicas across different geographic locations. It relies on PBFT~\cite{castro99practical} as agreement protocol and uses HMAC-SHA-256 as MACs to authenticate the messages exchanged between replicas. (2)~\textbf{\hft} employs a hierarchical system architecture running the two-level Steward protocol~\cite{amir10steward} to coordinate multiple sites that each host a dedicated cluster of replicas. Steward requires threshold cryptography for which \hft uses the scheme proposed by Shoup~\cite{shoup00practical} based on 1024-bit RSA signatures. (3)~\textbf{\system} represents our system architecture proposed in this paper. In this evaluation, \system's agreement group runs PBFT for consensus and its \channel{}s protect their messages with 1024-bit RSA signatures.


\begin{figure}[b!]
	\includegraphics[page=1]{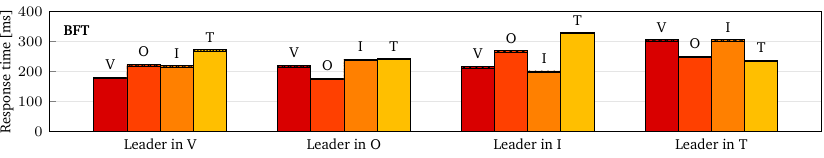}%
	\vspace{-2mm}\\
	\includegraphics[page=2]{eval-leader.pdf}%
	\vspace{-2mm}\\
	\includegraphics[page=3]{eval-leader.pdf}%
	\caption{50th~(\raisebox{-.3mm}{\protect\tikz \protect\node[draw, minimum width=7, minimum height=7] {};}) and 90th~(\raisebox{-.3mm}{\protect\tikz \protect\node[draw, minimum width=7, minimum height=7, postaction={pattern=crosshatch, pattern color=black}] {};}) percentiles of write latencies for different client and leader locations including Virginia~(V), Oregon~(O), Ireland~(I), and Tokyo~(T).}
	\label{fig:eval-leader-all}
\end{figure}


To conduct experiments in an actual wide-area environment, we start virtual machines~(t3.small, 2\,VCPUs, 2\,GB\,RAM, Ubuntu\,18.04.4\,LTS, OpenJDK\,11) in 4~Amazon EC2 regions across the globe (Virginia, Oregon, Ireland, and Tokyo). In each of these regions, we deploy 50~clients that issue 100 writes/reads per second (200~bytes) to a key-value store provided by our systems under test; client messages carry 1024-bit RSA signatures. Given this client setting, our architectures demand the following replica placement for~$f=1$: For \bft, one replica is hosted in each of the 4~regions. \hft expects a cluster of 4~replicas in each region, which is used as contact cluster for local clients. For \system, we deploy one execution group (3~replicas) per region, distributed across different availability \linebreak zones. In addition, we start \system's 4~agreement replicas in separate Virginia availability zones.

\subsection{Writes}
%
In our first experiment, we examine the latency of writes issued by clients at different sites. Based on the results presented in Figure~\ref{fig:eval-leader-all}, we make three important observations: (1)~In all evaluated architectures the response times to a major degree depend on a client's geographic location. For \bft and \hft, clients in Virginia for example benefit from the fact that their local replica~(cluster) experiences comparably short round-trip times when communicating with its counterparts in Oregon and Ireland. In particular, this results in low latency when the Virginia replica (cluster) acts as leader of the wide-area consensus protocol and is able to reach a quorum together with these two other sites. In \system, clients in Virginia also observe low write latency, but for a different reason. Here, the fact that the agreement group resides in the same region as the clients' local execution group allows clients in Virginia to achieve response times of as low as 13~milliseconds. (2)~For each client location, \system \mbox{provides} significantly lower latency than \bft~(up to 95\,\%) and \hft~(up~to~94\,\%). This is a direct consequence of the fact that in contrast to the other two system architectures \system does not execute a full-fledged replication protocol over wide-area links. Instead, a write request only has to wait for two wide-area hops: from a client's local execution group to the agreement group and back. The distribution of the ordered write request to other execution groups is handled by the agreement group and thus does not require execution groups to explicitly wait for each other. That is, when an execution replica in \system receives an \textsc{Execute} for a write from the agreement group, the replica can immediately process the operation and return a reply to the client. (3)~The response times of \bft and \hft vary considerably depending on the position of the current leader of the wide-area consensus protocol. \hft clients in Ireland, for example, experience a 53\,\% higher latency when the leader is positioned in Tokyo compared to when the leader role is assigned to Virginia. In contrast, the specific location of the agreement-group leader in \system only has a negligible effect on overall response times due to all agreement replicas residing in the same region, resulting in stable response times even across leader changes.


\begin{figure}[b!]
	\vspace{-2mm}
	\subfloat[Strongly consistent reads\label{fig:eval-readonly-strong}] {
		\includegraphics[clip, trim=0 1mm 0 0,page=1]{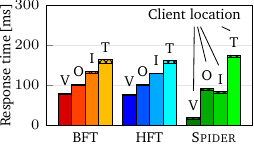}
	}
	\hfill
	\subfloat[Weakly consistent reads\label{fig:eval-readonly-weak}] {
		\includegraphics[clip, trim=0 1mm 0 0,page=2]{eval-readonly.pdf}
	}
	\hfill
	\subfloat[Writes\label{fig:eval-modularity}] {
		\includegraphics[clip, trim=0 1mm 0 0]{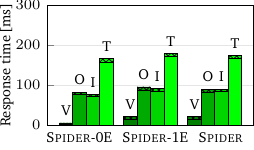}
	}
	\vspace{-3mm}
	\caption{50th~(\raisebox{-.3mm}{\protect\tikz \protect\node[draw, minimum width=7, minimum height=7] {};}) and 90th~(\raisebox{-.3mm}{\protect\tikz \protect\node[draw, minimum width=7, minimum height=7, postaction={pattern=crosshatch, pattern color=black}] {};}) percentiles of reads with different consistency guarantees and writes in different \system variants. Client and leader locations are Virginia~(V), Oregon~(O), Ireland~(I), and Tokyo~(T).}
	\label{fig:eval-readonly}
\end{figure}

\subsection{Reads}
%
In our second experiment, we compare the evaluated architectures regarding their (fast-)paths for reads with different consistency guarantees. As Figure~\ref{fig:eval-readonly-strong} shows, response times of strongly consistent reads in \system display a similar pattern as writes due to following the same path through the system. For clients in Tokyo, this leads to slightly higher response times compared with \bft and \hft, which in this case benefit from directly querying replicas without intermediaries in between. For all other client locations, \system's approach, which only requires waiting for one wide-area round trip from a client's execution group to the agreement group and back, enables lower latency than provided by \bft and \hft. With regard to weakly consistent reads~(see Figure~\ref{fig:eval-readonly-weak}), both \hft and \system achieve response times of 2~milliseconds or less, as these operations can be entirely handled by replicas in a client's vicinity and thus do not require wide-area communication as~in~\bft.

\subsection{Modularity Impact}
%
In our third experiment, we quantify the impact of our decision to design \system as a modular architecture that separates agreement from execution and consists of loosely coupled replica groups connected via \channel{}s. We create two variants of \system where (1)~the agreement group also executes requests and is the only group in the system~(\systemze) and (2)~there is only one execution group that is co-located with the agreement group in Virginia~(\systemoe). While, \systemze allows us to study \system without \channel and externalized execution, based on \systemoe we can assess the influence of an \channel without wide-area delays. Our results show that when clients access \systemze and \systemoe from different sites, response times are dominated by the wide-area communication between clients and replicas. Thus, the modularization overhead is small and adds less than 14~milliseconds~(see Figure~\ref{fig:eval-modularity}).


\begin{figure}[b!]
	\subfloat[Writes\label{fig:eval-newloc-avg-write}] {
		\includegraphics[page=1]{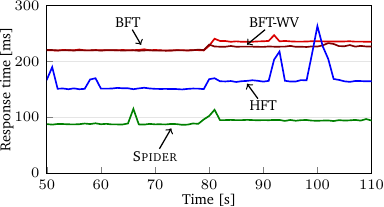}
	}
	\hspace{2mm}
	\subfloat[Weakly consistent reads\label{fig:eval-newloc-avg-read}] {
		\includegraphics[page=2]{eval-newloc-avg.pdf}
	}
	\caption{Impact of a new client site on overall latency.}
	\label{fig:eval-newloc-avgs}
\end{figure}

\subsection{Adaptability}
%
In our fourth experiment, we evaluate the write and read performance new clients experience when they join the system at an additional location. For this purpose, we start with our usual setting and after 80~seconds launch 50~clients in the EC2 region S\~{a}o Paulo. Once running, the new clients in \bft and \hft issue their requests to existing replicas, while for \system they contact an additional execution group also set up in S\~{a}o Paulo. Involving more client sites than replica sites in \bft and \hft, the setting in this experiment represents a typical use-case scenario for weighted-voting approaches~(see Section~\ref{sec:background}). We therefore repeat the experiment with a fourth system~(\bftwv) that extends \bft with weighted voting and comprises a replica at each of the five client locations. As required by weighted voting, two of the five replicas are assigned higher weights in the consensus protocol. Specifically, these are the replicas in Virginia and Oregon because this weight distribution achieves the best performance in our evaluation scenario. Figure~\ref{fig:eval-newloc-avgs} presents the results of this experiment showing the average response times observed across all active client sites. We omit the results for strongly consistent reads as they show a similar picture as writes. For each system, we evaluate different leader locations, but for clarity Figure~\ref{fig:eval-newloc-avgs} only reports the results of the configurations achieving the lowest response times for each system.

Figure~\ref{fig:eval-newloc-avg-write} shows that the overall write latency increases for all evaluated architectures once the clients in S\~{a}o Paulo join the system. This is a consequence of the fact that due to its geographic location EC2's S\~{a}o Paulo region has comparably high transmission times to other cloud regions. Clients in S\~{a}o Paulo therefore observe response times between about 124~milliseconds (\system) and about 298~milliseconds (\bft), which alone causes the measurable jumps in the overall write latency averages; the response times for clients in other regions remain unaffected.
Interestingly, \bft and \bftwv achieve similar write performance throughout the experiment and thereby confirm that weighted voting does not automatically improve response times. This is only true when the additional replica is located at a site that is better connected than the existing ones and therefore enables the wide-area consensus protocol to reach faster quorums. In the setting evaluated here, \bft's typical consensus quorum is based on votes of the replicas in Virginia, Oregon, and Ireland and therefore already provides better performance than any combination that includes S\~{a}o Paulo.

As shown in Figure~\ref{fig:eval-newloc-avg-read}, of the evaluated architectures \system is the only one that allows the new clients in S\~{a}o Paulo to perform weakly consistent reads with low latency. While all other systems require the clients in S\~{a}o Paulo to read from at least one remote replica and consequently experience overall read-latency increases of up to 23~milliseconds, \system makes it possible to introduce an execution group in the new region to efficiently handle the reads of local clients.


\begin{figure}[b]
	\includegraphics{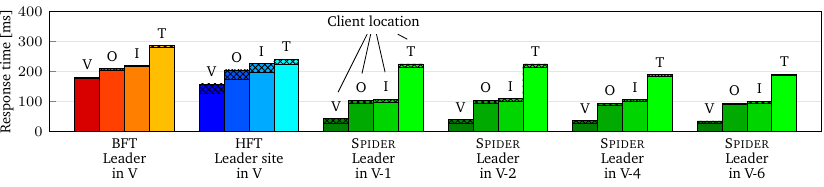}%
	\vspace{-0.5mm}
	\caption{50th~(\raisebox{-.3mm}{\protect\tikz \protect\node[draw, minimum width=7, minimum height=7] {};}) and 90th~(\raisebox{-.3mm}{\protect\tikz \protect\node[draw, minimum width=7, minimum height=7, postaction={pattern=crosshatch, pattern color=black}] {};}) percentiles of write latencies for system configurations tolerating $f=2$ faults.}
	\label{fig:eval-leader-all-f2}
\end{figure}

\subsection{Tolerating Two Faults}
%
In our fifth experiment, we examine write latencies for settings that are configured to tolerate $f=2$ faults in each agreement and execution group.
We place the additional replicas into nearby EC2 regions (Ohio, California, London, Seoul) to make use of further fault domains.
The results in Figure~\ref{fig:eval-leader-all-f2} show that due to increased communication latency within groups both \hft and \system see a moderate increase of response times by up to 46 milliseconds compared with the $f=1$~setting, with \system still providing significantly lower latency than \bft and \hft.

\subsection{\channel{} Implementations}
\label{sec:evaluation-channels}

In our final set of experiments, we evaluate the two \channel variants presented in Section~\ref{sec:implementations}. For this purpose, in addition to the AWS-based setting used in previous experiments, we also include a local testbed that emulates the Amazon EC2 environment and allows us to conduct an extensive evaluation with higher throughputs. In our local testbed, each server~(Intel Xeon E3-1275 v5 CPU, 4 physical cores with two hyper-threads, 16\,GB RAM, Ubuntu 20.04.4 LTS, OpenJDK~11) represents a different cloud region and hosts all clients and replicas of that region. To emulate the wide-area latencies between regions, we measure the corresponding latencies in AWS and apply them using \texttt{netem}~\cite{hemminger05network}, a Linux feature which delays packets sent via Ethernet. For the communication within a region, we inject one-way delays of 0.2~milliseconds between processes. All servers of our testbed have 1~Gbit Ethernet interfaces and are connected to the same switch.

\begin{figure}
	\subfloat[Throughput\label{fig:eval-channels-lat}] {
		\includegraphics[page=2]{eval-channels-msig.pdf}
	}\\
	\subfloat[CPU usage\label{fig:eval-channels-cpu}] {
		\includegraphics[page=1]{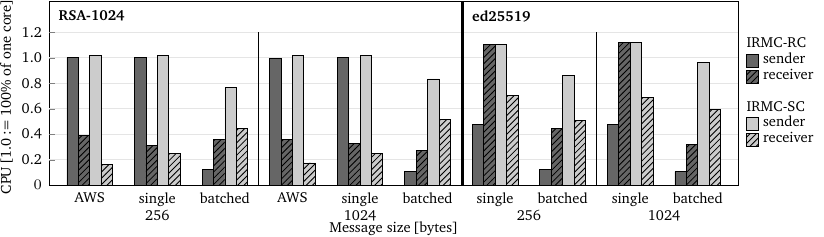}
	}\\
	\subfloat[Network usage\label{fig:eval-channels-net}] {
		\includegraphics[page=3]{eval-channels-msig.pdf}
	}\\
	\subfloat[Network overhead (i.e., ratio of actual bytes sent to payload-message size)\label{fig:eval-channels-net-overhead}] {
		\includegraphics[page=4]{eval-channels-msig.pdf}
	}
	\caption{Performance and resource usage of different \channel implementations for different signature types.}
	\label{fig:eval-channels}
\end{figure}

\headline{Configurations}
%
Figure~\ref{fig:eval-channels} shows the measurement results for a scenario in which an \channel connects the agreement group in Virginia with an execution group in Tokyo. Numbers obtained in the public cloud are labeled ``AWS'', while ``single'' marks the equivalent run in the local testbed. In addition, ``batched'' refers to a testbed configuration in which the signature batching optimization is enabled~(see Section~\ref{sec:signature-batching}). For each configuration, we evaluate both channel variants~(\channela and \channelb) submitting messages of different sizes~(256~bytes and 1024~bytes). Furthermore, we assess the impact of the authentication scheme by differentiating between a traditional signature~(RSA-1024) and a modern signature based on elliptic curves~(ed25519). Besides throughput, we report the measured CPU and network usage as well as network overhead; the latter denotes the ratio of bytes actually sent over the network to the number of bytes that comprise the payload. For the sender side, the presented CPU usage originates from the replica that serves as \channelb sender, whereas the receiver-side consumption is taken from the replica that processes the largest number of messages. The reported data-transfer numbers are the aggregate amount of transmitted data.

\headline{\channel Variants}
%
The evaluation results confirm the two \channel implementations to have individual characteristics. Without the need to verify signatures for \textsc{Certificate} messages, when signature batching is deactivated \channela sender endpoints require less CPU resources per message than \channelb senders and therefore enable \channela{}s to achieve a higher maximum throughput~(e.g.,~3,400 vs. 2,479~requests per second in the AWS setting). On the other hand, due to forwarding only one wide-area message per receiver endpoint, \channelb{}s significantly reduce the amount of data transferred over long-distance links~(e.g.,~for 256 byte messages the measurement results show a reduction by a factor of 3.0), thereby saving costs in public-cloud environments.

\headline{RSA-1024}
%
Authenticating messages with RSA-1024, the throughput of the unoptimized~(i.e.,~non-batching) \channel variants is limited by the signature computation saturating the CPU at the sender side. Under such conditions, it is possible to drastically increase the achievable throughput of an \channel by applying the signature batching optimization described in Section~\ref{sec:signature-batching}. Specifically, in our local testbed we observed improvements by up to a factor of 5.1 for the \channela and a factor of up to 8.5 for the \channelb. The CPU usage at the sender side falls to 13\%~(\channela) and 83\%~(\channelb), respectively, showing that the benchmark is no longer CPU-bound. \channelb's higher sender-side CPU usage compared with \channela is a result of the need to exchange and verify signatures while assembling certificates. The fact that \channelb achieves a higher throughput than \channela in this case is a result of \channela incurring a higher wide-area data transmission overhead. In particular, \channela's throughput for 1024 byte messages is limited by the network, which according to \texttt{iperf3} in our testbed can transmit at most 111~MiB/s between regions.


\headline{ed25519}
%
Relying on ed25519 signatures, the throughputs of the unoptimized \channel{}s are lower than with RSA-1024~(e.g., about 2,600 vs. 4,200~requests per second for \channela). Comparing the two variants, \channela's throughput is about 50\% higher than that of \channelb, resulting in a larger performance difference than observed for RSA-1024. For ed25519, the CPU usage at the \channela receiver side significantly increases and becomes the bottleneck. This is a result of ed25519 signatures being computationally more expensive to verify than to create. In contrast, for \channelb the receiver-side CPU usage increases to only about 70\%. Here, the sender side reaches its CPU limit before the receiver side due to having to verify signatures while assembling a certificate.

Applying signature batching, the throughput increases by up to a factor of 8.8 for \channela and a factor of 16.0 for \channelb. In this scenario, CPU usage is no longer the bottleneck and therefore the throughput is able to reach a similar level as with RSA-1024. The increase in throughput comes at the cost of a higher network overhead, which is especially noticeable for \channelb. Here, the local traffic overhead increases more than three times compared with the non-batching variant. The reason for this is that a batch signature contains the path through the Merkle tree in addition to the signature itself. At the configured maximum batch size of 128~requests, this results in the inclusion of up to seven SHA-256 hashes~(i.e.,~at most 224~bytes in total). Overall, the ed25519~network overhead is smaller than the overhead associated with RSA-1024 due to the use of smaller signatures.

\headline{Summary}
%
Overall, our results show that signature batching is an effective technique to increase the throughput of \channel{}s. We believe further improvements to be possible by minimizing the optimization's network overhead. Specifically, sender endpoints can be implemented to transmit the entire Merkle tree only once and afterwards simply refer to it in later messages.

\section{Related Work}
\label{sec:related}

\headline{Adaptive BFT Replication}
%
\system is not the first work to argue that it is crucial to enable BFT systems to dynamically adapt to changing conditions. Abstract~\cite{aublin15next} makes it possible to substitute the consensus protocol at runtime, for example, switching to a more robust algorithm once a replica failure has been suspected or detected. CheapBFT~\cite{kapitza12cheapbft} and ReBFT~\cite{distler16resource} follow a similar idea by comprising two different agreement protocols~(one for the normal case and one for fault handling) of which only one is active at a time. In contrast, the reconfiguration mechanism developed by Carvalho et al.~\cite{carvalho18dynamic} for BFT-SMaRt~\cite{bessani14state} temporarily runs two consensus algorithms in parallel to achieve a more efficient switch. As a result of \system's modularity, integrating support for the dynamic substitution of the agreement protocol is feasible and the use of customized protocols designed for high \linebreak performance~\cite{martin06fast,behl15consensus} or strong resilience~\cite{amir10prime,aublin13rbft} would not require modifications to execution groups.

Other works allow BFT systems to dynamically change specific protocol properties. Depending on the current workload, de S\'{a} et al.~\cite{desa13adaptive}, for example, vary the parameters deciding how many requests are batched together and ordered within a single consensus instance. Berger~et~al.~\cite{berger19resilient} rely on a weighted voting scheme~\cite{sousa15separating} and by changing weights adjust the individual impact a replica has on the agreement outcome. Adapting the batch size can be a measure to improve the performance of \system's agreement group.
In contrast, the use of a weighted voting scheme in general is only effective if (1)~a system contains more than the minimum number of agreement replicas and (2)~agreement \linebreak replicas are located in different geographic regions; both of these points do not apply to \system.

\headline{Communication Between Replica Groups}
%
Amir et al. proposed BLinks~\cite{amir07customizable} as a means to send the totally ordered outputs of one replicated state machine to another replicated state machine that uses them as inputs. Unfortunately, the requirement of a channel-wide total order prevents \system from relying on BLinks as execution replicas do not necessarily have to use the same order when submitting new requests to the agreement group via their request channels. \channel{}s, on the other hand, do not have this restriction and furthermore comprise a built-in flow-control mechanism that represents the basis of \system's global flow control. However, transmitting only a single message between one dedicated sender and one dedicated receiver, BLinks may be used as a template for an \channel{} implementation that involves even fewer wide-area messages than \channelb.

\headline{Partitioned Agreement Groups}
%
GeoBFT~\cite{gupta20resilientdb} uses replica groups in different regions that each run a full agreement protocol.
In each protocol round every group orders a request yielding a request certificate, which is shared with all other groups.
Afterwards the requests are merged into a single total order and are executed.
This requires all groups to distribute a certificate in every round, even if it just contains a placeholder request, and thus all groups must work at the same time to make progress.
In \system this requirement only applies to the agreement group whereas a limited number of slow execution groups can be skipped.
Sharing a request ordering certificate in GeoBFT works by having the leader replica forward it to $f+1$~replicas of each group, which then further forward the certificate within their group.
This request distribution scheme represents a middle ground between BLinks and \channelb{}s.
Unlike \channel{}s it is coupled with the agreement protocol and has to remotely trigger a \linebreak view-change to replace a leader that does not complete the request distribution in a timely manner.

\headline{Efficient Client Communication}
%
In most BFT systems, clients need to receive replies from different replicas in order to prove a result correct~\cite{castro99practical}, which in geo-replicated settings can significantly increase the number of messages exchanged over wide-area links. SBFT~\cite{gueta19sbft} addresses this problem by adding a protocol phase that aggregates request acknowledgments of multiple replicas into a single message to the client. In Troxy~\cite{li18troxy}, a client also has to wait for a single reply only, because the reply voter is hosted inside a trusted domain at the server side and forwards its decisions to the client through a secure channel. In \system, clients are typically located in the same region as an execution group allowing for communication over short-distance links. For scenarios in which this is not the case, it would be possible to extend \system to use one of the approaches discussed above.

\headline{Leader Selection in Geo-Replicated Systems}
%
Multiple authors have underlined the impact that the leader-replica location has on response times, independent of the fault model, and presented solutions to select the leader in a way that minimizes overall latency~\mbox{\cite{sousa15separating,liu17leader,eischer18latency}}. Other agreement-based systems do not need to determine a fixed leader as they continuously rotate the leader role among replicas~\cite{veronese09spin,mao09towards,veronese10ebawa,mao08mencius,milosevic13bounded}. As our experiments show, with agreement replicas residing in different availability zones of the same cloud region, the specific location of the consensus leader in \system only has a negligible effect on response times. Consequently, \system achieves low and stable latency without requiring means to dynamically select or rotate the leader.

\headline{Crash-tolerant Wide-Area Replication}
%
Several works addressed the efficiency of geo-replication in systems that unlike \system solely tolerate crashes, not Byzantine failures. In Pileus~\cite{terry13consistency}, for example, writes are only handled by a subset of replicas that first order and execute them, and then bring all other replicas up to date by transferring state changes. P-Store~\cite{schiper10pstore} improves efficiency in wide-area environments by performing partial replication, thereby freeing a site from the need to receive and process all updates. Clock-RSM~\cite{du14clock} establishes a total order on requests by exploiting the timestamps of physical clocks and without requiring a dedicated leader replica. EPaxos~\cite{moraru13there} in contrast does not rely on a total request order, but only orders those requests that interfere with each other due to accessing the same state parts.

\section{Conclusion}

The cloud-based \system system architecture models a BFT system as a collection of loosely coupled replica groups that can be flexibly distributed in geo-replicated environments. In contrast to existing approaches, \system does not require the execution of complex multi-phase protocols over wide-area links, but instead performs essential tasks such as consensus, leader election, and checkpointing across replicas residing in the same region. Our experiments show that this approach enables \system to achieve low and stable response times.


\bibliographystyle{ACM-Reference-Format}
\bibliography{paper}

\end{document}